\documentclass[showpacs,pra,aps,twocolumn]{revtex4}
\usepackage{bm}
\usepackage{amsmath}
\usepackage{amssymb}
\usepackage{latexsym}
\usepackage{amsfonts}
\usepackage{epsfig}
\usepackage{psfrag}
\usepackage{subfigure}
\usepackage{color}

\begin{document}

\title{Localization of weakly interacting Bose gas in quasiperiodic potential}
\author{Sayak Ray, Mohit Pandey, Anandamohan Ghosh and S. Sinha}
\affiliation{Indian Institute of Science Education and Research-Kolkata, Mohanpur, Nadia-741246, India}

\date{\today}

\begin{abstract}
We study the localization properties of weakly interacting Bose gas in a quasiperiodic potential commonly known as Aubry-Andr\'e model.  Effect of interaction on localization is investigated by computing the `superfluid fraction' and `inverse participation ratio'. For interacting Bosons the inverse participation ratio increases very slowly after the localization transition due to `multisite localization' of the wave function. We also study the localization in Aubry-Andr\'e model using an alternative approach of classical dynamical map, where the localization is manifested by chaotic classical dynamics.  For weakly interacting Bose gas, Bogoliubov quasiparticle spectrum and condensate fraction are calculated in order to study the loss of coherence with increasing disorder strength. Finally we discuss the effect of trapping potential on localization of matter wave. 

\end{abstract}

\pacs{67.85.-d, 05.30.Jp, 03.75.Hh}

\maketitle

\section{Introduction}
In recent years `Anderson localization' of particles and waves has regained interest in quantum many particle physics. 
`Anderson localization' is a remarkable quantum phenomenon for which the propagating wave becomes exponentially localized in the presence of disorder\cite{anderson,rmp} and is a well studied subject in the context of electronic systems in presence of disorder. Like particles, waves can also localize in disordered medium.
Localization of `matter wave' has recently been observed in experiments on ultracold Bose gases in presence of speckle potential\cite{aspect} and in bichromatic optical lattices\cite{inguscio1}. For certain parameters, Bosons in bichromatic lattice can be mapped on to Aubry-Andr\'e (AA) model\cite{aa} with `quasi-periodic' potential. Also localization of light in AA model has been observed experimentally\cite{lahini}. 
Unlike Anderson model with random disorder, the AA model exhibits localization transition in one dimension at a critical strength of the potential. Apart from that the quasi-periodicity of AA model gives rise to various interesting spectral properties\cite{pichard}. In recent experiment \cite{inguscio2} effect of interaction on AA-model has been studied.  
Repulsive interaction plays an important role in the formation of correlated phases like `Bose-glass' phase\cite{BHM,lugan,inguscio3,pasienski}. 
Also the dynamics and diffusion of interacting Bose gas in presence of disorder have been investigated both experimentally\cite{inguscio4,clement} and theoretically\cite{larcher,flach}.
In recent years `many body localization'\cite{huse1} and localization at finite temperature\cite{huse2,alt,shlyapnikov} have generated an impetus to study disordered Bose gas.

In this work we investigate localization of weakly interacting Bose gas in AA potential. Apart from calculating the ground state properties, we also study localization of wavefunction by using a classical dynamical map. 
The paper is organized as follows: In section II, we discuss non-interacting Bose gas in quasiperiodic potential. In subsection A, we review the AA model and discuss the single particle localization properties. Localization in non-interacting system using dynamical map approach is presented in subsection B.
In section III, various physical quantities like inverse participation ratio and superfluid fraction are computed to investigate localization of weakly interacting Bose gas within mean-field theory. Effect of non-linearity due to interactions on the dynamical map is studied. Further, we compute the Bogoliubov quasiparticle energies and amplitudes to investigate quantum fluctuations. We study the localization of both Bogoliubov amplitudes and non-condensate densities.

% We also compute the Bogoliubov quasiparticle energies and amplitudes to investigate both quantum fluctuations induced by localization and localization of non-condensate densities.
 Effect of trapping potential on localization and effect of disorder on the center of mass motion are presented in section IV. Finally we summarize our results in section V. 

\section{Non-interacting Bosons in quasiperiodic potential}
In the original experiment\cite{inguscio1} localization of ultracold Bosons has been studied using bichromatic optical lattice which can be mapped on to Aubry-Andr\'e model within tight binding approximation and for certain parameter regime\cite{modugno2}.
The Aubry-Andr\'e model is defined by the Hamiltonian,
\begin{equation}
H= - J\sum_n   (|n\rangle \langle n + 1|+ |n+1\rangle \langle n|) + \lambda \sum_n \cos(2\pi \beta n) |n\rangle \langle n|
\label{aah}
\end{equation}
where $|n\rangle$ is a Wannier state at lattice site $n$, $J$ is nearest neighbour hopping strength, $\lambda$ is the strength of onsite potential and period of potential is determined by $\beta$. In the rest of the paper, we would be working in the units in which $J=1$. 

For $\lambda=2$, the Hamiltonian \eqref{aah} is equivalent to the Harper model\cite{harper} describing the motion of an electron in a square lattice in the presence of a perpendicular magnetic field, where the flux $\Phi$ through each plaquette in units of flux quantum $\Phi_{0}= h/e$ is given by the parameter $\beta = \Phi/\Phi_{0}$. Energy spectrum of this problem gives rise to the well known `Hofstadter butterfly'\cite{hofstadter}.
The Hamiltonian given in Eq.\eqref{aah}, poses very interesting properties for irrational values of $\beta$. When $\beta$ is chosen to be a `Diophantine number', AA model undergoes a localization transition at a critical value of the potential strength $\lambda = 2$\cite{aa,jit1}. On the contrary, localization transition is absent in one dimensional Anderson model with random disorder. The quasiperiodic potential generates a correlated disorder in the AA model. 

\subsection{Single particle localization}
In order to study localization transition in AA model, we choose $\beta= (\sqrt{5}-1)/2$ which is the inverse of `golden mean' and a Diophantine number . This is particularly helpful since rational approximation of $\beta$ can be done by Fibonacci series. Although incommensurability of the potential with the underlying lattice plays an crucial role in AA model, for numerical studies one can approximate $\beta = F_{n-1}/F_{n}$, where $F_{n}$ is $n$th Fibonacci number for sufficiently large value $n$. This rational approximation of $\beta$ fixes the lattice size $N_{s}= F_{n}$ to impose periodic boundary condition\cite{hanggi}. 

\begin{figure}
%\centering
%\rotatebox{0}
%\begin{center}
\includegraphics[width=8cm]{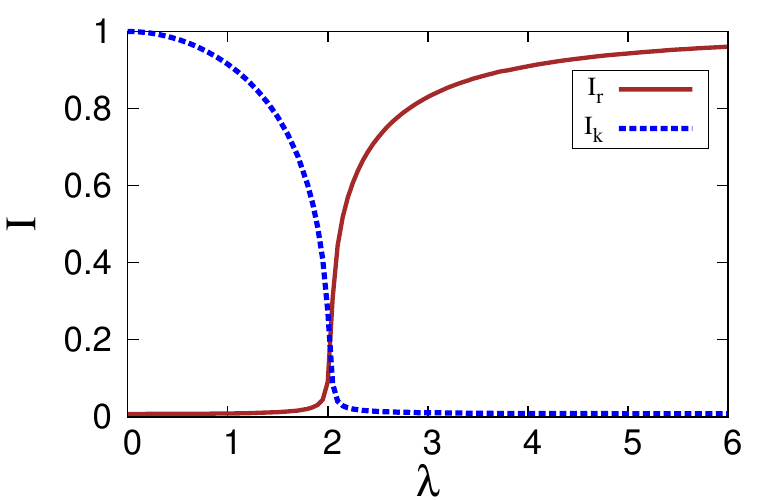}
%\end{center}
\caption{IPR of ground state wavefunction as a function of strength of the potential $\lambda$. IPR in real space $I_{r}$ and in dual momentum space $I_{k}$ are shown by solid(black) line and dashed (blue) line respectively.}
\label{fig1}
\end{figure}

Duality in AA model can be shown by introducing new basis states in momentum space,
\begin{equation}
|k\rangle  = N_{s}^{-1/2} \sum_n \exp(i 2\pi k\beta n)| n \rangle.
\label{dual_basis}
\end{equation}
The dual model is obtained by substituting Eq.\ref{dual_basis} in Eq.\ref{aah},
\begin{equation}
H= \frac{\lambda}{2} \left[ \sum_k  |k\rangle \langle k + 1|+ \mbox{h.c}  + \frac{4}{\lambda} \sum_k \cos(2\pi \beta k) |k\rangle \langle k| \right].
\label{dual_aa}
\end{equation} 
It is important to note that AA model becomes self-dual at a critical coupling $\lambda = 2$, where the localization transition occurs. 
To obtain the eigenvalues and eigenfunctions of the Hamiltonian we expand the state vector in terms of Wannier states, $|\psi\rangle = \sum_{n} \psi_{n}|n\rangle$, where $\psi_{n}$ is the wavefunction at $n$th lattice site. The eigenvalue equation of the Hamiltonian Eq.\ref{aah} is reduced to a discrete Schr\"odinger equation,
\begin{equation}
-(\psi_{n+1} + \psi_{n-1}) + \lambda \cos(2\pi \beta n) \psi_{n} = \epsilon \psi_{n},
\label{dse}
\end{equation}
where $\epsilon$ is the energy eigenvalue. The degree of localization of a normalized state $|\psi\rangle$ can be quantified by `inverse participation ratio' (IPR) $I$,
\begin{equation}
I = \sum_{n}|\psi_{n}|^{4}
\label{ipr_eq}
\end{equation}
The wavefunction of an extremely localized particle at a site $n_{0}$ is given by $\psi_{n} = \delta_{n,n_{0}}$, for which the IPR becomes unity. On the other hand, for the completely delocalized wavefunction $\psi_{n} = 1/\sqrt{N_{s}}$, IPR is $1/N_{s}$ which vanishes in the thermodynamic limit. For AA model all energy eigenfunctions in real space are exponentially localized and IPR sharply increases to unity above the critical coupling $\lambda = 2$. Due to the duality of AA model, the localization of wavefunction in real space and in dual momentum space shows opposite behavior. In real space the wavefunctions are localized above $\lambda =2$, wheras localization in dual momentum space occurs for $\lambda \leq 2$. The IPR of the ground state wavefunction in real space and in dual monentum space as a function of  $\lambda$ are shown in Fig.\ref{fig1}. It is interesting to note that IPR in real and momentum space intersects at the self dual point $\lambda = 2$.
%%%%%%%%%%%%%%%%%%%%%%%%%%%%%%%%%%%%%%%
%%%%%%%%%%%%%%%%%%%%%%%%%%%%%%%%%%%%%%%%%%

\begin{figure}
%\rotatebox{0}
%\begin{center}
\subfigure[]{\includegraphics[width=4.2cm]{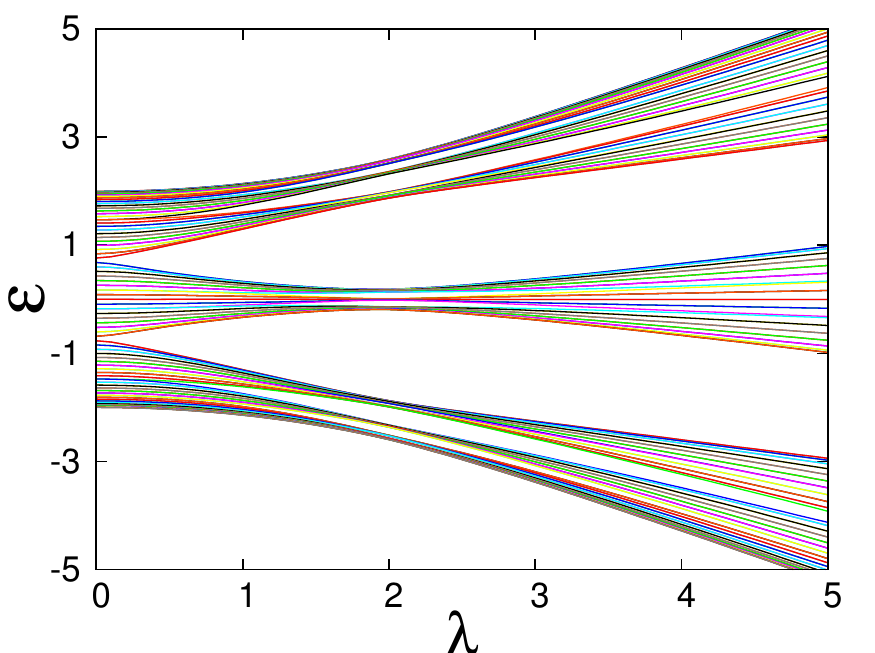}}
\subfigure[]{\includegraphics[width=4.2cm]{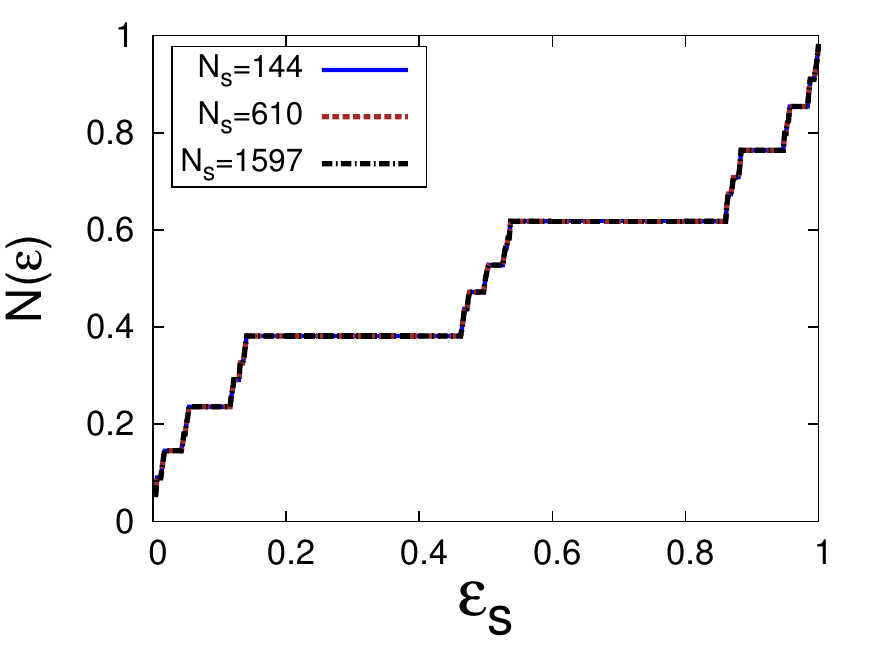}}
%\end{center}
\caption{(a) Eigenenergies of AA model as a function of $\lambda$ for system size $N_{s}=144$. (b) Devil's staircase structure of the integrated density of states $N(\epsilon)$ scaled by $N(\epsilon_{max})$ as a function of scaled energy $\epsilon_{s} = (\epsilon - \epsilon_{min})/(\epsilon_{max} -\epsilon_{min})$ at $\lambda =2$ and for different system sizes. Here $\epsilon_{min}$ and $\epsilon_{max}$ are the minimum and maximum energy eigenvalues.}
\label{fig2}
\end{figure}

The spectrum of AA model also shows various interesting properties. For $\lambda =0$ it has usual single band energy spectrum of periodic lattice. Addition of the potential generates quasiperiodic structure and destroys the band like dispersion. Successive rational approximation of $\beta$ generates more periodicity over the underlying lattice, which in turn breaks the original energy band into many subbands and it leads to opening of band gaps. The energy spectrum of AA model is obtained by numerical diagonalization and is depicted in Fig.\ref{fig2}a for increasing values of $\lambda$. The variation of energy gaps with the coupling strength can be noticed in this figure. The spectral statistics and distribution of energy gaps are analyzed in \cite{pichard}. 
Mathematically it can be shown that the energy spectrum of AA model forms a Cantor set\cite{jit}. Self-similarity in energy levels can be understood from the integrated level density,
\begin{equation}
N(\epsilon) = \sum_{i}\theta(\epsilon - \epsilon_{i}),
\label{int_ds}
\end{equation}
where $\epsilon_{i}$ is the i'th eigenvalue and $\theta(x)$ is heaviside step function. The integrated density of states $N(\epsilon)$ shows `Devil's staircase' like fractal structure, which is evident from Fig.\ref{fig2}b where we plotted the normalized integrated density of states $N(\epsilon)$ as a function of scaled energy within the interval of zero to one. 
For different values of $\lambda$, the Devil's staircase structures of $N(\epsilon)$ do not overlap, which indicates that the fractal dimension changes with the strength of the potential $\lambda$.
%Although the integrated density of states exhibits Devil's staircase like 
%fractal structure, its fractal dimension changes with the strength of the 
%potential $\lambda$.
%For this reason the rescaled values of $N(\epsilon$ do not overlap on each 
%other when they are compared for different values of $\lambda$.
%%%%%%%%%%%%%%%%%%%%%%%%%%%%%%%%%%%%%%%
%%%%%%%%%%%%%%%%%%%%%%%%%%%%%%%%%%%%%%%%%%

\begin{figure}
\centering
%\rotatebox{0}
%\begin{center}
\includegraphics[width=8cm]{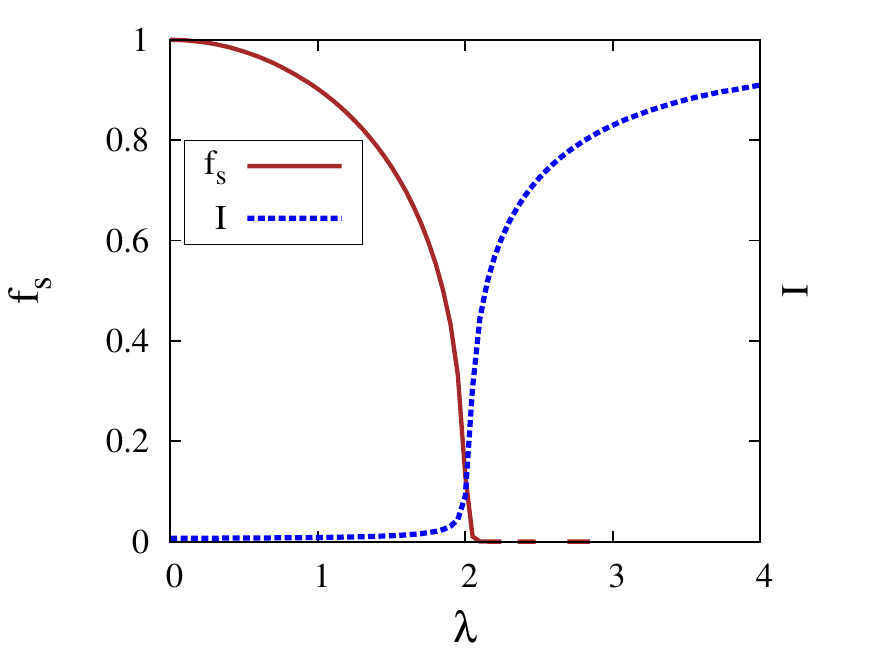}
%\end{center}
\caption{Superfluid fraction $f_{s}$ (solid line) and IPR of ground state in real space $I_{r}$ (dashed line) as a function of $\lambda$.}
\label{fig3}
\end{figure}

Transport properties also show significant changes in localization transition. Due to spatial localization of single particle states transport coefficients vanish in the  thermodynamic limit. In electronic systems the conductivity vanishes exponentially with the system size due to `Anderson localization'. For neutral superfluid corresponding physical quantity is `superfluid fraction'(SFF), which is measured by generating a superflow by applying a phase twist\cite{fisher}. In presence of the phase twist the original Hamiltonian in Eq.\ref{aah} becomes,
\begin{equation}
H_{\Theta}  =  -\sum_{n=1}^{N_s} (e^{- i \Theta/ N_s} |n\rangle \langle n + 1|+ \mbox{h.c}) +  \lambda \cos(2\pi \beta n) |n\rangle \langle n|,
\label{aa_twist}
\end{equation}
where $\Theta$ is arbitrarily small phase difference across the boundary. For one dimensional system with periodic boundary condition this is exactly similar to a quantum ring in presence of a flux which generates supercurrent through the ring.
The superfluid fraction $f_{s}$ is defined as\cite{burnett},
\begin{equation}
f_s = N_{s}^{2}  \frac{E_{0}(\Theta)-E_{0}(0)}{\Theta^2},
\label{sf}
\end{equation}
where $E_{0}(\Theta)$ is the ground state energy of the Hamiltonian $H_{\Theta}$ with arbitrary small value of $\Theta$, and $E_{0}(0)$ is the ground state energy of the original Hamiltonian given in Eq.\ref{aah}. To obtain the ground state energy upto $\Theta^2$ order, the Hamiltonian $H_{\Theta}$ is expanded as,
\begin{equation}
H_{\Theta} \approx H + \frac{\Theta}{N_s} \hat{J} - \frac{\Theta^2}{2 N_s^2}\hat{T}
\label{pert}
\end{equation}
where we define a current operator $\hat{J} =   -i \sum_n (|n\rangle \langle n + 1| -  \mbox{h.c}) $ and the usual kinetic energy $\hat{T} =-\sum_n (|n\rangle \langle n + 1|+  \mbox{h.c})$. 
Using second order perturbation, the SFF in Eqn.\ref{sf} can be written as,
\begin{equation}
f_{s} = -\frac{1}{2} \langle{\psi_0}|{\hat{T}}|{\psi_0}\rangle + \sum_{i \neq 0} \frac{{|\langle \psi_{i}|}{\hat{J}}{|\psi_0}\rangle|^2}{\epsilon_{i}-\epsilon_{0}},
\label{sf_pert}
\end{equation}
where $\epsilon_{i}$, $|\psi_{i}\rangle$ are eigenvalues and eigenfunctions of Hamiltonian given in Eq.\ref{aah} and $|\psi_{0}\rangle$ is ground state wavefunction. 
Variation of SFF with the coupling strength of the potential is shown in Fig.\ref{fig3}.  For increasing values of $\lambda$ the SFF decreases from unity and vanishes at the critical point $\lambda = 2 $.
%%%%%%%%%%%%%%%%%%%%%%%%%%%%%%%%%%%%%%%
%%%%%%%%%%%%%%%%%%%%%%%%%%%%%%%%%%%%%%%%%%

\subsection{Hamiltonian map}
Localization phenomena can also be studied by an alternative method of classical Hamiltonian map(CHM), which is very useful to obtain analytical estimate of localization length\cite{izrailev_rev,izrailev}. The discrete Schr\"odinger equation given in Eq.\ref{dse} can be written as following dynamical map,
\begin{eqnarray}
p_{n+1} & = & p_{n} + (\lambda \cos(2\pi \beta n) -\epsilon -2)x_{n},\\
x_{n+1} & = & x_{n} + p_{n+1}
\label{class_eqn}
\end{eqnarray}
where the classical dynamical variables are given by, $x_{n} = \psi_{n}$ and $p_{n} = \psi_{n} - \psi_{n-1}$. Here the wavefunction $\psi_{n}$ plays the role of position in CHM and the number of lattice site becomes the number of iteration or time axis of the dynamics. We can choose initial real wavefunctions $\psi_{0}$ and $\psi_{1}$ (or equivalently $x_{1}$ and $p_{1}$) and evolve the dynamical variables by the transfer matrix,
\begin{equation}
T_{n} = \left[\begin{array}{c c}  (\lambda \cos(2\pi \beta n) -\epsilon -1) & 1\\
(\lambda \cos(2\pi \beta n) -\epsilon -2) & 1
 \end{array} \right].
\label{trans_mat}
\end{equation}
%%%%%%%%%%%%%%%%%%%%%%%%%%%%%%%%%%%%%%%%%%%%%%%%%%%%%%%%%%%%%%%%%%%%%%%%%%%%%%%%%
\begin{figure}[ht]
\subfigure[]{\includegraphics[scale=0.12]{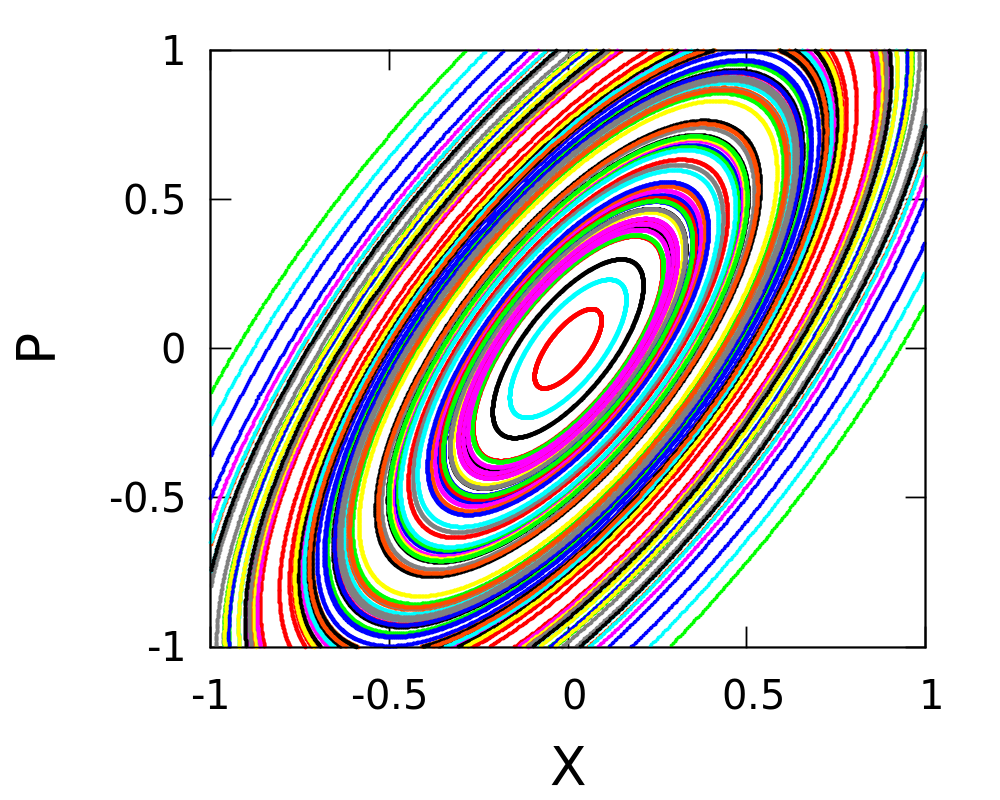}}
\subfigure[]{\includegraphics[scale=0.12]{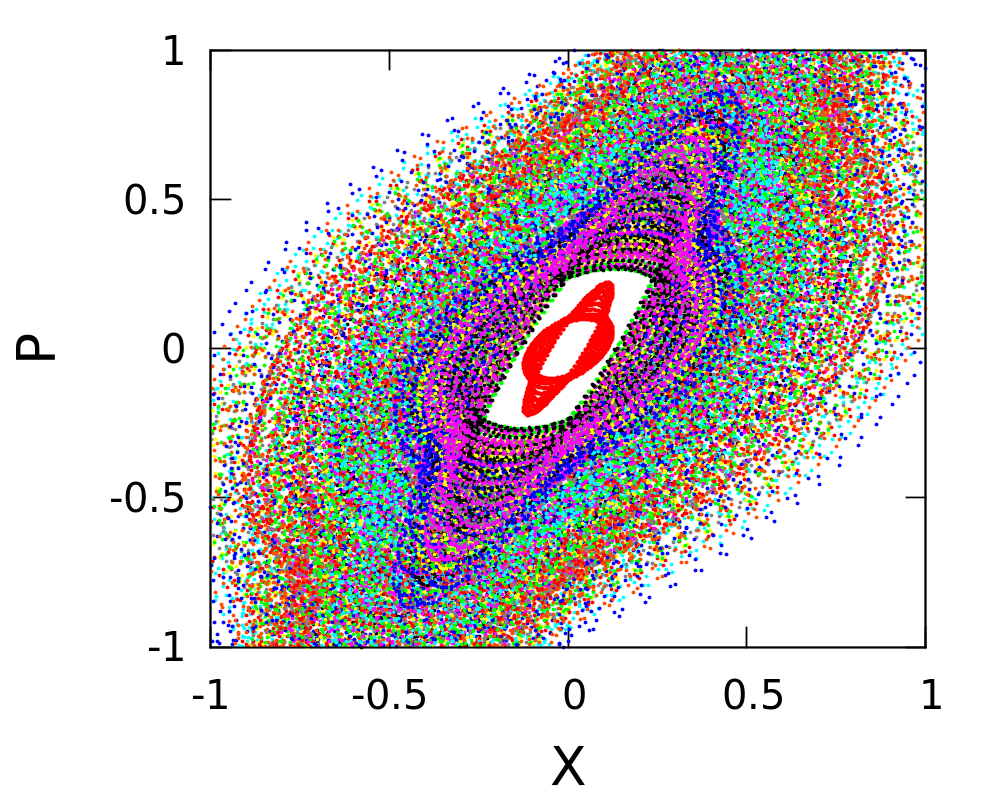}}
\subfigure[]{\includegraphics[scale=0.12]{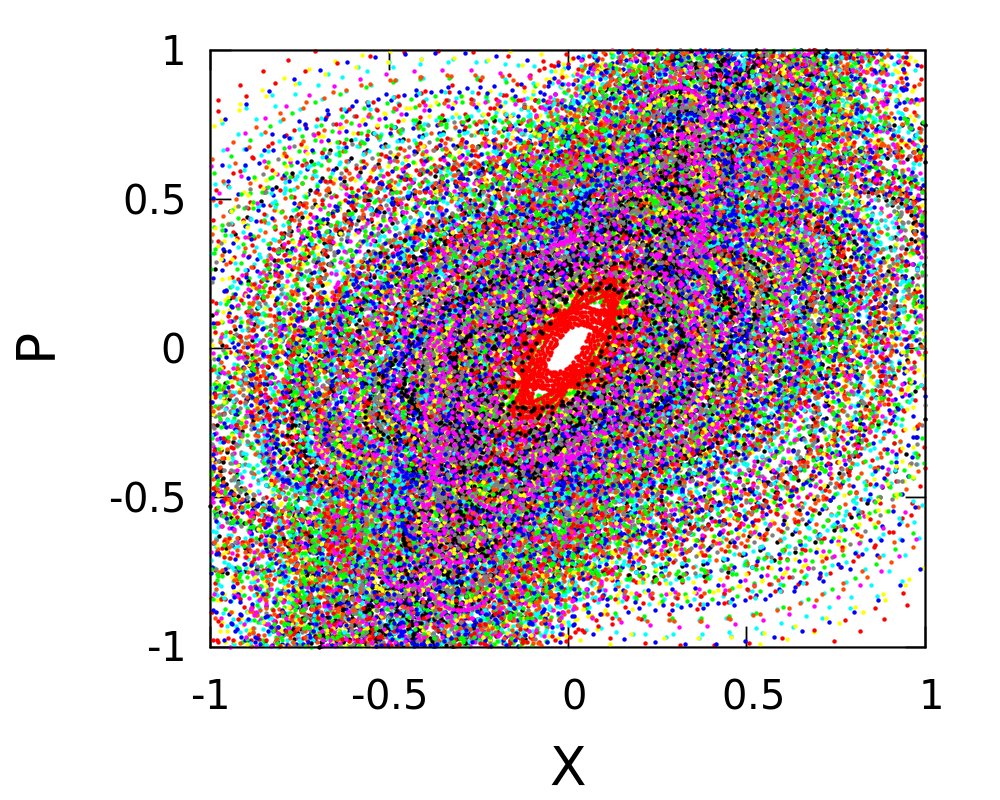}}
\subfigure[]{\includegraphics[scale=0.12]{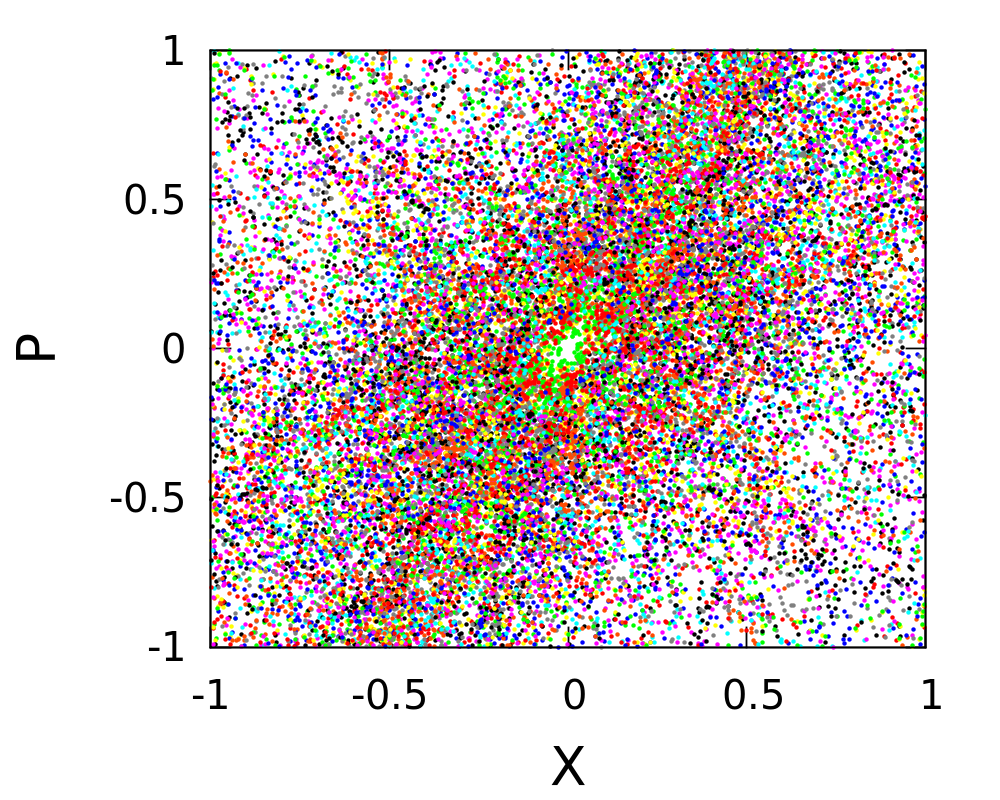}}
\caption{Phase portraits of CHM corresponding to Eq.\ref{class_eqn} with a fixed energy eigenvalue close to the middle of the energy band and for an ensemble of initial conditions. Fig.(a) to (d) corresponds to $\lambda$=0, 0.6, 1.2, 1.8 respectively.}
\label{fig4}
\end{figure}
%%%%%%%%%%%%%%%%%%%%%%%%%%%%%%%%%%%%%%%%%%%%%%%%%%%%%%%%%%%%%%%%%%%%%%%%%%%%%%%%%
By iteration of this map we obtain the asymptotic behavior of the dynamical variables for a fixed energy eigenvalue $\epsilon$. Mathematically it can be shown that the dynamical instability of the above Hamiltonian map corresponds to the localization phenomena. Exponentially localized wavefunctions asymptotically fall off as $\psi_{n} \sim e^{-n/\xi}$, where $\xi$ is the localization length. But for arbitrary choice of the initial wavefunction, the solution of Eq.\ref{dse} also has an exponentially growing component $\sim e^{n/\xi}$. Hence the localization is manifested by the exponential growth of the dynamical variables of the Hamiltonian map which gives rise to chaos in the classical phase space. Whereas the extended states represents periodic motion in the phase space. To understand the localization transition in AA model, we choose an eigenvalue close to the band center and calculate the phase space trajectories of the Hamiltonian map  for an ensemble of random initial values of $x$ and $p$. The phase portraits of Eq.\ref{class_eqn} are shown in Fig.\ref{fig3} for increasing values of  $\lambda$. 
In the delocalized regime with small disorder strength $\lambda$, the phase portrait consists of closed elliptic orbitals as seen from Fig.3a.  Increasing the strength of quasiperiodic potential $\lambda$ leads to diffusive behavior at the outer part of the phase space and regular portion of phase space with periodic orbits is reduced (see Fig.3b and Fig.3c). Finally for $\lambda$ close to the critical value, an instability sets in the CHM and phase space trajectories in Fig.3d show chaotic behavior which indicates localization transition.

\begin{figure}[ht]
\includegraphics[scale=0.9]{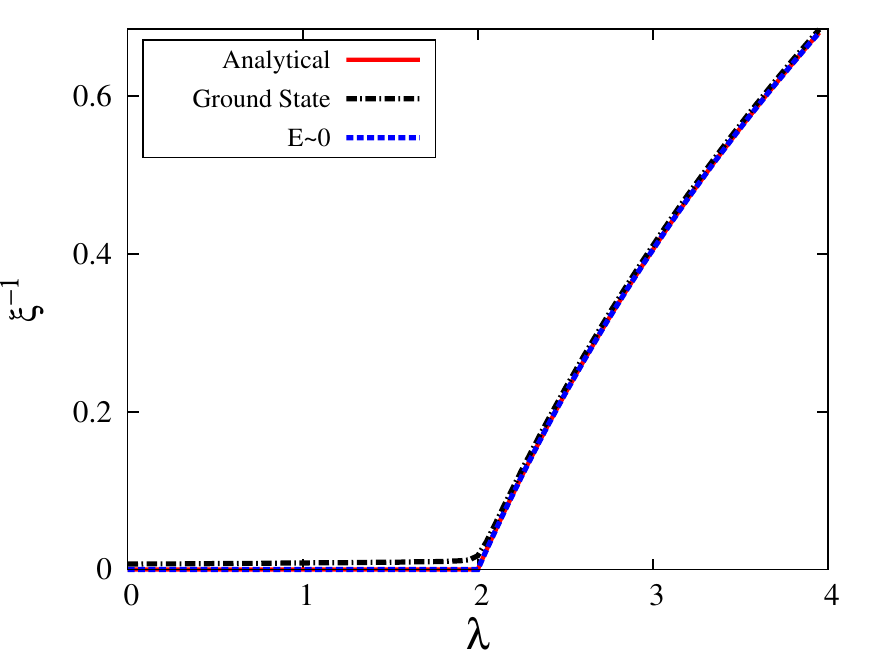}
\caption{Variation of Lyapunov exponent or inverse localization length $1/\xi$ with disorder strength $\lambda$. Dotted and dot-dashed lines represent numerically obtained Lyapunov exponents from Eq.\ref{le} and for comparison the analytical result is shown by solid (red) line.}
\label{fig5}
\end{figure}

In dynamical systems `Lyapunov exponent' (LE) is a measure to quantify chaos.
Since the number of lattice site plays the role of time in the classical map, the LE corresponds to the inverse localization length of the wavefunction. For periodic motion LE vanishes and in the chaotic regime the non-vanishing LE gives finite localization length $\xi$. To calculate the LE we construct the matrix $U_{N_s}= T_{N_s}T_{N_s-1}.....T_{2}T_{1}$ by multiplying the transfer matrices $T_{n}$ sequentially for $N_s$ iteration. The LE $l$ can be obtained by using the formula,
\begin{equation}
l = \lim_{N_s\rightarrow \infty} \frac{\log(\lambda_{+})}{N_s},
\label{le}
\end{equation}
where $\lambda_{+}$ is the largest eigenvalue of the matrix $U_{N_s}$ and LE is obtained from $\xi = 1/l$. The localization length can also be calculated by using Thouless formula \cite{thouless}. Using the duality of AA model the analytical expression of LE is given by $\xi = \log(\lambda/2)$\cite{aa}, which is independent of energy. Within CHM approach we numerically compute the localization length $\xi$ from Eq.\ref{le} using QR decomposition method \cite{lauter}. 
Numerically obtained localization length as a function of $\lambda$ for different energy eigenvalues are compared with the analytical result in Fig.\ref{fig5}. 
%%%%%%%%%%%%%%%%%%%%%%%%%%%%%%%%%%%%%%%%%%%%%%%%%%%%%%%%%%%%%%%%%%%%%
%%%%%%%%%%%%%%%%%%%%%%%%%%%%%%%%%%%%%%%%%%%%%%%%%%%%%%%%%%%%%%%%%%%%%

\begin{figure}[ht]
\subfigure[]{\includegraphics[scale=0.48]{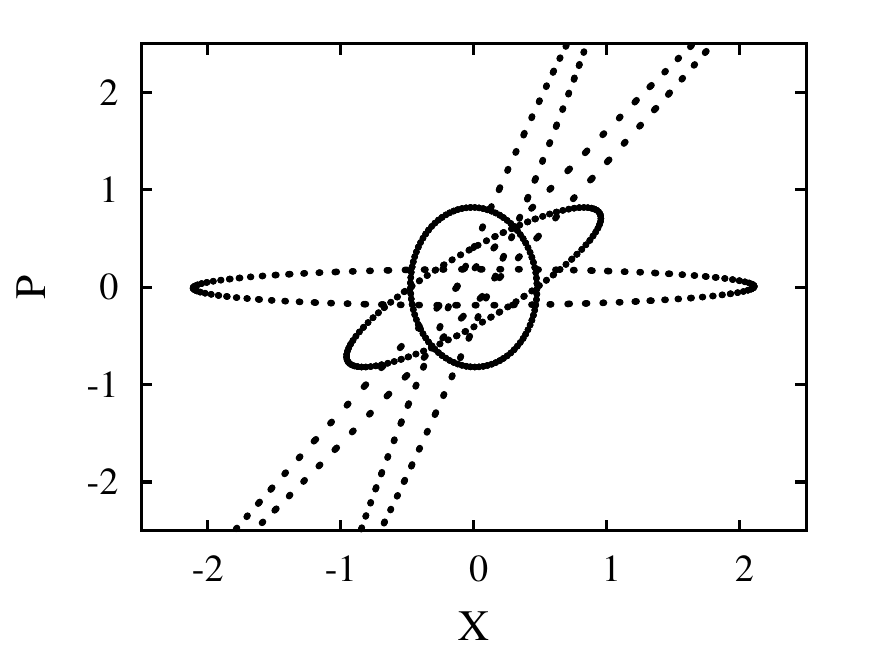}}
\subfigure[]{\includegraphics[scale=0.48]{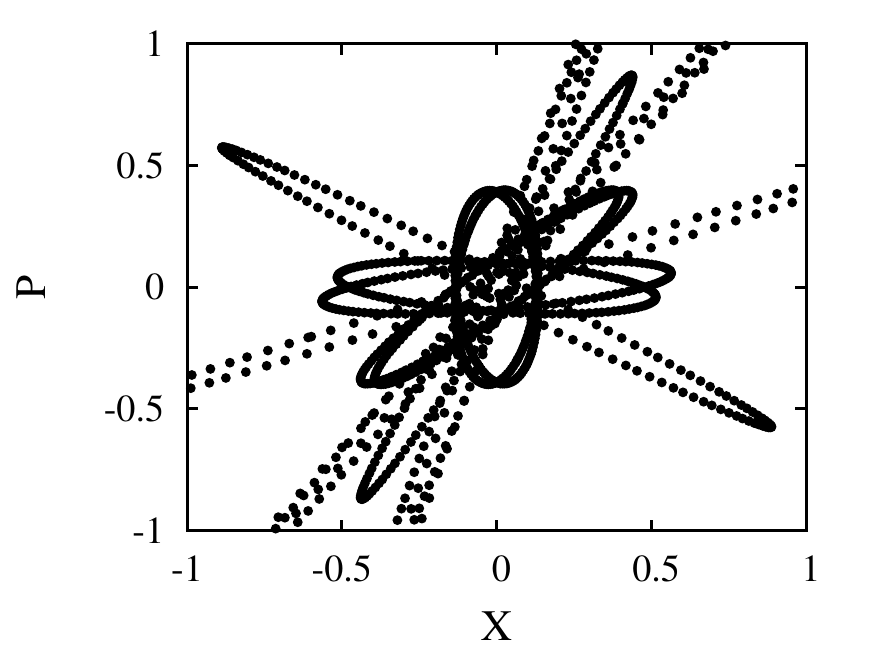}}
\subfigure[]{\includegraphics[scale=0.48]{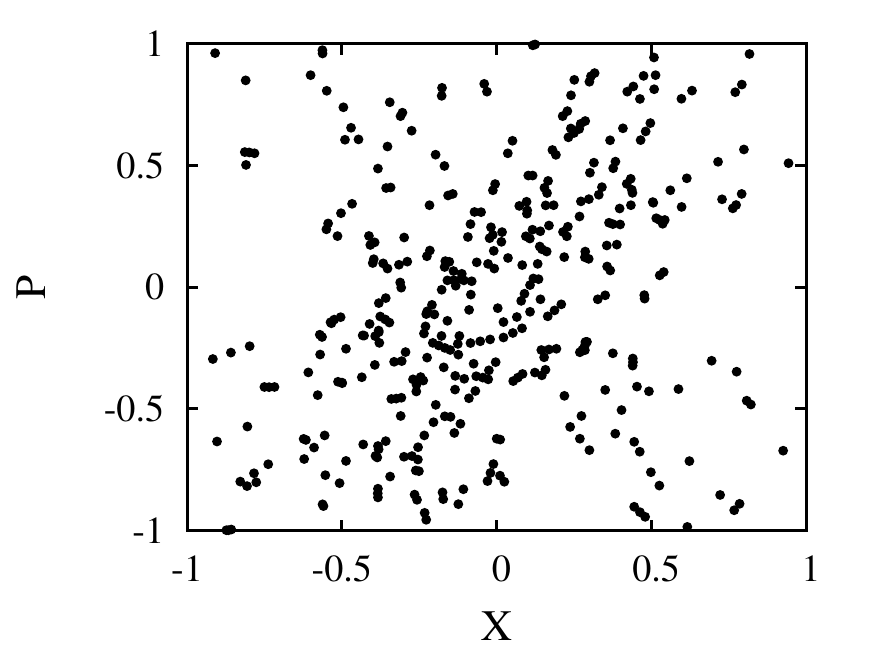}}
\subfigure[]{\includegraphics[scale=0.48]{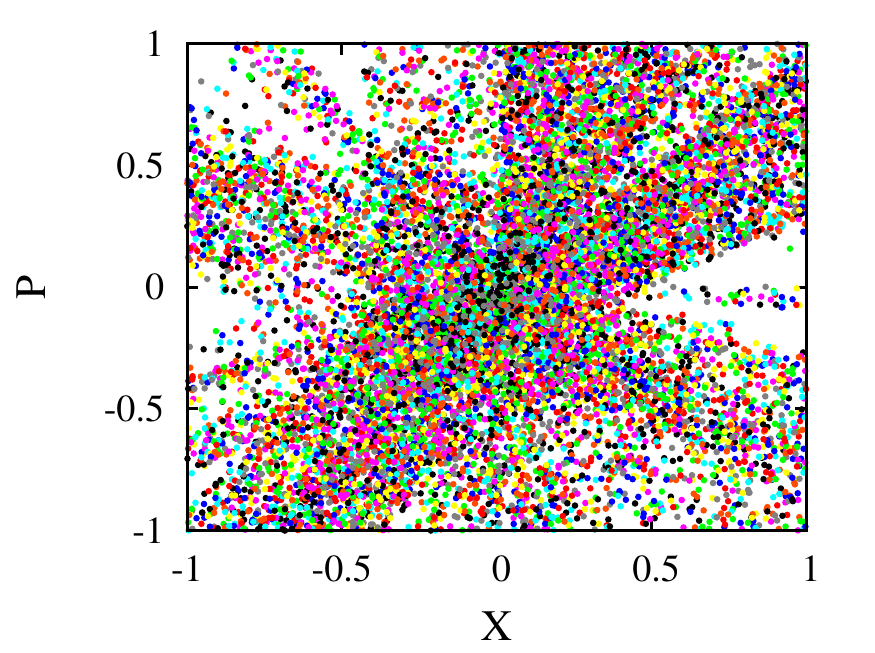}}
\caption{Phase space trajectories with successive rational approximation of $\beta$ and for $\lambda=2.2$. Here (a) to (c) corresponds  to $\beta _5=0.6$, $\beta _7=0.615$, $\beta _{12}=0.618$ respectively with same initial conditions and (d) corresponds to $\beta _{14}=0.618$ with an ensemble of initial conditions.}
\label{fig6}
\end{figure}

For the localization transition in AA model with quasiperiodic potential, the choice of $\beta$ as irrational number (particularly a Diophantine number) plays a crucial role \cite{jit1}. To understand this mathematical condition, we analyze the phase space of CHM for successive rational approximation of $\beta$, which is shown in Fig.\ref{fig6}. Using Fibonacci series, the rational approximation of $\beta$ can be written as $\beta_{n} = F_{n-1}/F_{n}$ and the potential has periodicity $F_{n}$ for successive integer $n$. For sufficiently large value of $n$, $\beta_{n}$ approaches to the inverse of `golden mean' and the potential becomes quasiperiodic. 
In the localized regime, for $\lambda=2.2$ the phase space trajectories of CHM with increasing order of rational approximation of $\beta$ are presented in Fig.\ref{fig6} (a) to (c) for fixed initial condition. Even in the localized regime phase space contains periodic orbits for $\beta_{5}$. As shown in Fig.6b to Fig.6d, the periodic orbits break as $\beta$ approaches to the inverse of `golden mean' by successive rational approximation and finally dynamics become chaotic which is consistent with the localization phenomena.  
%%%%%%%%%%%%%%%%%%%%%%%%%%%%%%%%%%%%%%%%%%%%%%%%%%%%%%%%%%%%%%%%%%%%%%%%%%%%%%%%%%  
%%%%%%%%%%%%%%%%%%%%%%%%%%%%%%%%%%%%%%%%%%%%%%%%%%%%%%%%%%%%%%%%%%%%%%%%%%%%%%%%%% 

\section{Localization of weakly interacting Bose gas}
Interacting Bosons in presence of quasiperiodic potential can be described by Bose-Hubbard model \cite{BHM},
\begin{equation}
H = -\sum_{\langle ij\rangle}(a^{\dagger}_{i}a_{j} + \mbox{h.c}) + \lambda \sum_{i}\cos(2\pi \beta i) \hat{n}_{i} + \frac{U}{2}\sum_{i} \hat{n}_{i}(\hat{n}_{i}-1)
\label{Bose_hubbard}
\end{equation}
where $a^{\dagger}_{i}$($a_{i}$) are creation (annihilation) operators for Bosons at site $i$, $\hat{n}_{i} = a^{\dagger}_{i}a_{i}$, and $U$ is the strength of onsite repulsive interaction. Above quantum many-body Hamiltonian can capture various correlated phases of strongly interacting Bosons which undergoes quantum phase transition. It is known that Bose-Hubbard model in the presence of random disorder can give rise to `Bose glass' phase \cite{BHM}. 
For sufficiently weak interaction strength $U$ and for high average density of Bosons a `quasi-condensate' may form in one dimensional systems \cite{1d_shlyap}. In this regime, one can replace the bosonic operators $a_{i}$ by a classical field $\psi_{i}$ which represents the macroscopic wavefunction of the `quasi-condensate'. Minimization of the classical energy corresponding to Eq.\ref{Bose_hubbard} leads to the `discrete nonlinear Schr\"odinger equation' (DNLS),
\begin{equation}
-(\psi_{i+1} + \psi_{i-1}) +\lambda \cos(2\pi \beta i)\psi_{i}+ U|\psi_{i}|^{2} \psi_{i} = \mu \psi_{i},
\label{dnlse}
\end{equation}
where $\mu$ is the chemical potential, and the normalization of the wavefunction of $N_{b}$ number of Bosons gives $\sum_{i} |\psi_{i}|^{2} = N_{b}$. 
%%%%%%%%%%%%%%%%%%%%%%%%%%%%%%%%%%%%%%%%%%%%%%%%%%%%%%%%%%%%%%%%%%%%%%%%%%%%%%%%%%
\begin{figure}[ht]
\subfigure[]{\includegraphics[scale=0.8]{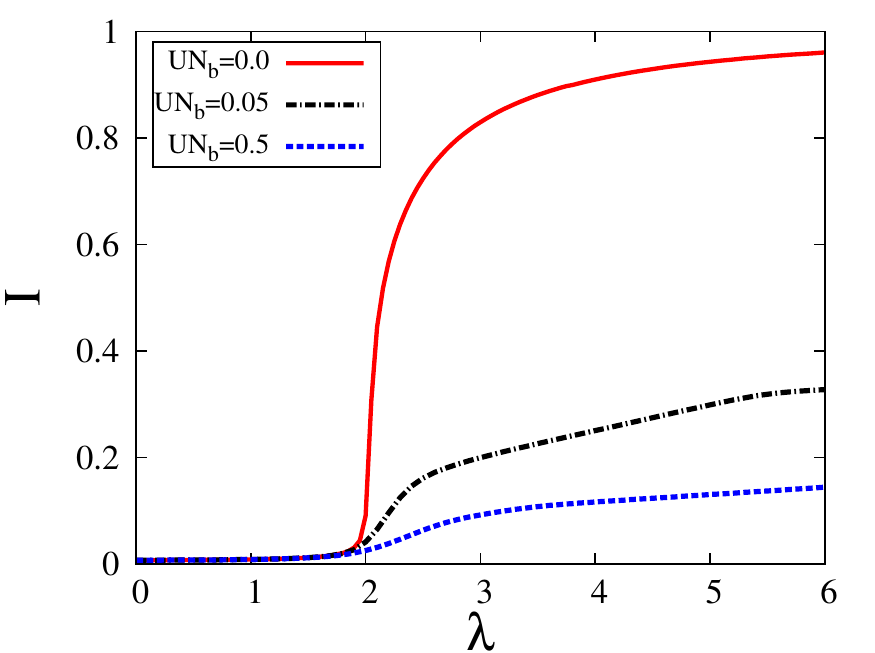}}
\subfigure[]{\includegraphics[scale=0.23]{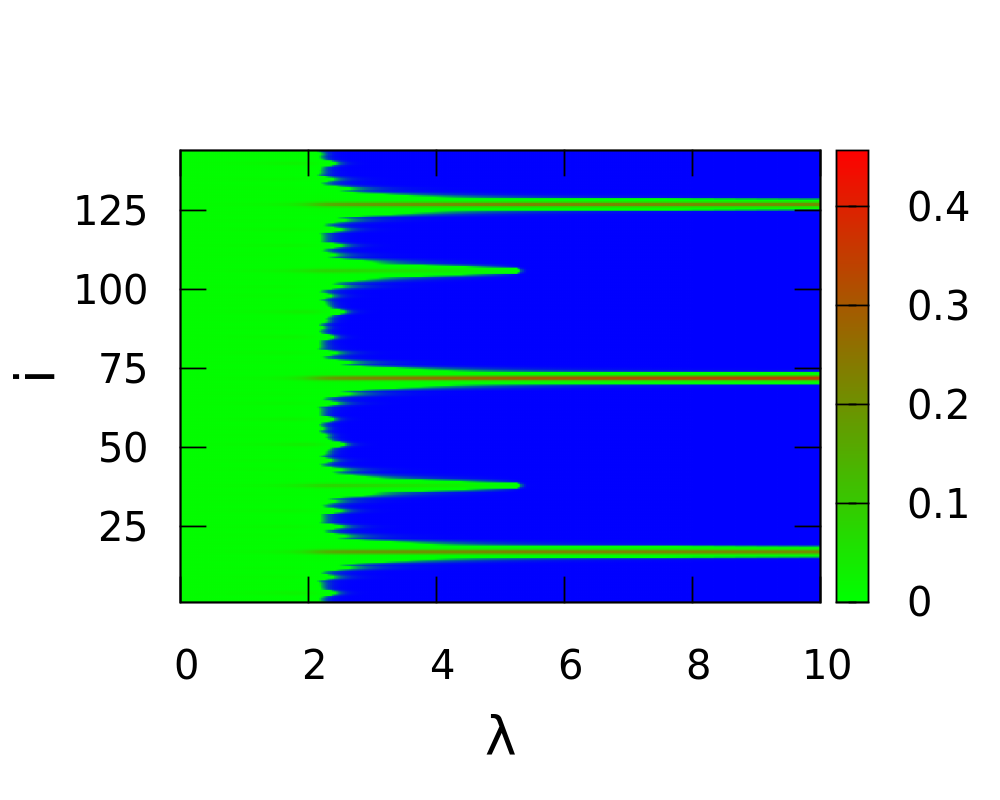}}
\caption{(a)Inverse Participation Ratio(I) of condensate wavefunction with increasing $\lambda$ for different $UN_b$. (b) Spatial variation of condensate density with $\lambda$ for $UN_b=0.05$, where \textit{i} denotes lattice site index.}
\label{fig7}
\end{figure}
%%%%%%%%%%%%%%%%%%%%%%%%%%%%%%%%%%%%%%%%%%%%%%%%%%%%%%%%%%%%%%%%%%%%%%%%%%%%%%%%%%

We numerically find out the ground state wavefunction of Eq.\ref{dnlse} by imaginary time propagation method. For weak disorder the wavefunction is extended and numerical convergence is fast, whereas close to the localization transition more care is needed to obtain the ground state since many metastable states appear in this region. To obtain the degree of localization of the ground state wavefunction in presence of repulsion we calculate the IPR in real space using Eq.\ref{ipr_eq}. The variation of IPR of condensate wavefunction with increasing strength of disorder $\lambda$ for different values of effective interaction $UN_{b}$ is shown in Fig.\ref{fig7}(a). The IPR becomes nonzero for $\lambda\ge 2$ and increases with a slower rate compared to the non-interacting system, indicating spreading of the wavefunction due to the repulsive interaction. The degree of localization decreases with increasing strength of repulsive interaction $UN_{b}$. In Fig.7(b) the spatial variation of the wavefunction with disorder strength $\lambda$ is represented by color scale plot. It is evident from this figure that single site localization is not favorable energetically and the wavefunction is localized at almost degenerate but spatially separated sites. This particular feature of the fragmented condensate has also been studied in \cite{Samuel}. Due to the multisite localization the IPR is much less than unity and increases very slowly with $\lambda$. A change in the slope of IPR with increasing $\lambda$ occurs when the number of localized sites decreases and we have checked that finally at very large value of $\lambda$ the wavefunction becomes localized at a single site. 

\begin{figure}[ht]
\includegraphics[scale=0.8]{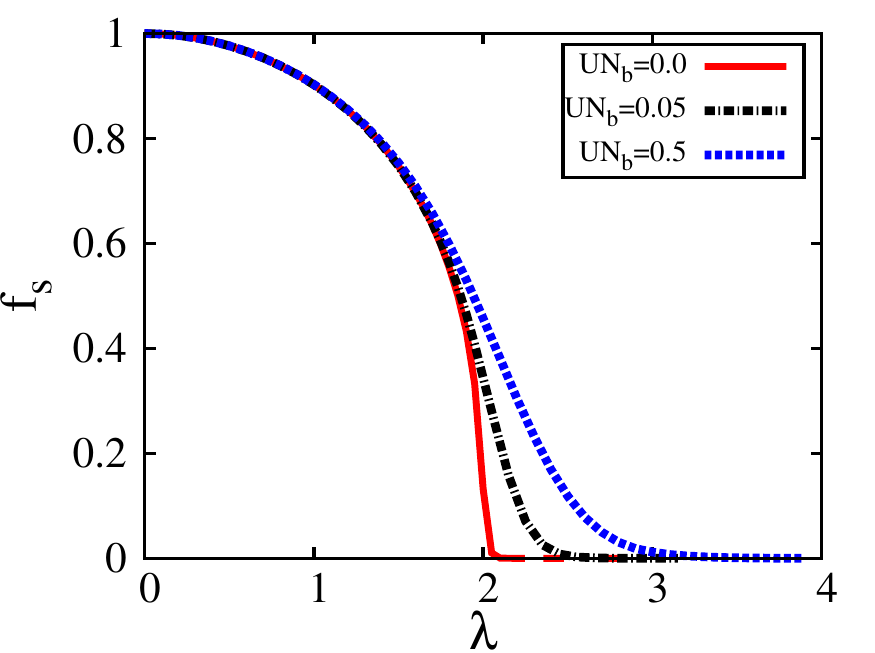}
\caption{Variation of `superfluid fraction' $f_s$ with  $\lambda$ for  different interaction strength $UN_b$.}
\label{fig8}
\end{figure}

To study the interplay between disorder and interaction in the transport properties of dilute Bose gas, we calculate the superfluid fraction $f_{s}$. To generate a superflow we introduce a small amount of phase twist ($\Theta \sim 0.1$) in the hopping term of DNLS (in Eq.\ref{dnlse}) similar to Eq.\ref{aa_twist} and then the superfluid fraction is computed using the formula,
\begin{equation} 
f_s = N_s ^2  \frac{E_{cl}(\Theta)-E_{cl}(0)}{\Theta^2},
\label{fs_int}
\end{equation}
where $E_{cl}(\Theta) $ is the classical energy corresponding to the Hamiltonian in Eq.\ref{Bose_hubbard},
\begin{eqnarray}
E_{cl}(\Theta) & =  & -\sum_{\langle ij\rangle}\left[e^{i\Theta/N_s}\phi^{\ast}_{i}\phi_{j} +\mbox{h.c}\right] + \lambda \sum_{i}\cos(2\pi \beta i) |\phi_{i}|^{2}\nonumber\\
& &  + \frac{UN_{b}}{2}\sum_{i} |\phi_{i}|^{4}.
\label{energy_gp}
\end{eqnarray}
The wavefunction $\phi_{i}$ minimize $E_{cl}(\Theta)$ and is normalized to unity. 
The superfluid fraction $f_{s}$ as a function of $\lambda$ for different values of repulsive interactions $UN_{b}$ is shown in Fig.\ref{fig8}. The SFF of weakly interacting Bose gas obtained from `density matrix renormalization group' also shows similar behavior \cite{minguzzi}.
Due to the repulsive interaction, $f_{s}$ vanishes at larger strength of quasiperiodic potential $\lambda > 2$, however the IPR rises from zero at $\lambda \approx 2$. This behavior is different from the noninteracting AA model. 
%%%%%%%%%%%%%%%%%%%%%%%%%%%%%%%%%%%%%%%%%%%%%%%%%%%%%%%%%%%%%%%%%%%%%%%%%%%%%%%
%%%%%%%%%%%%%%%%%%%%%%%%%%%%%%%%%%%%%%%%%%%%%%%%%%%%%%%%%%%%%%%%%%%%%%%%%%%%%%%

We also investigated the localization properties of DNLS by Hamiltonian map approach.
Eq.\ref{dnlse} can be written in the form of nonlinear classical map,
\begin{eqnarray}
p_{i+1} & = & p_{i} + (\lambda \cos(2\pi \beta i) -\mu -2)x_{i} + UN_{b}x_{i}^{3},\\
x_{i+1} & = & x_{i} + p_{i+1}
\label{class_dnls}
\end{eqnarray}
%%%%%%%%%%%%%%%%%%%%%%%%%%%%%%%%%%%%%%%%%%%%%%%%%%%%%%%%%%%%%%%%%%%%%%%%%%%%%%
\begin{figure}[ht]
\subfigure[]{\includegraphics[scale=0.12]{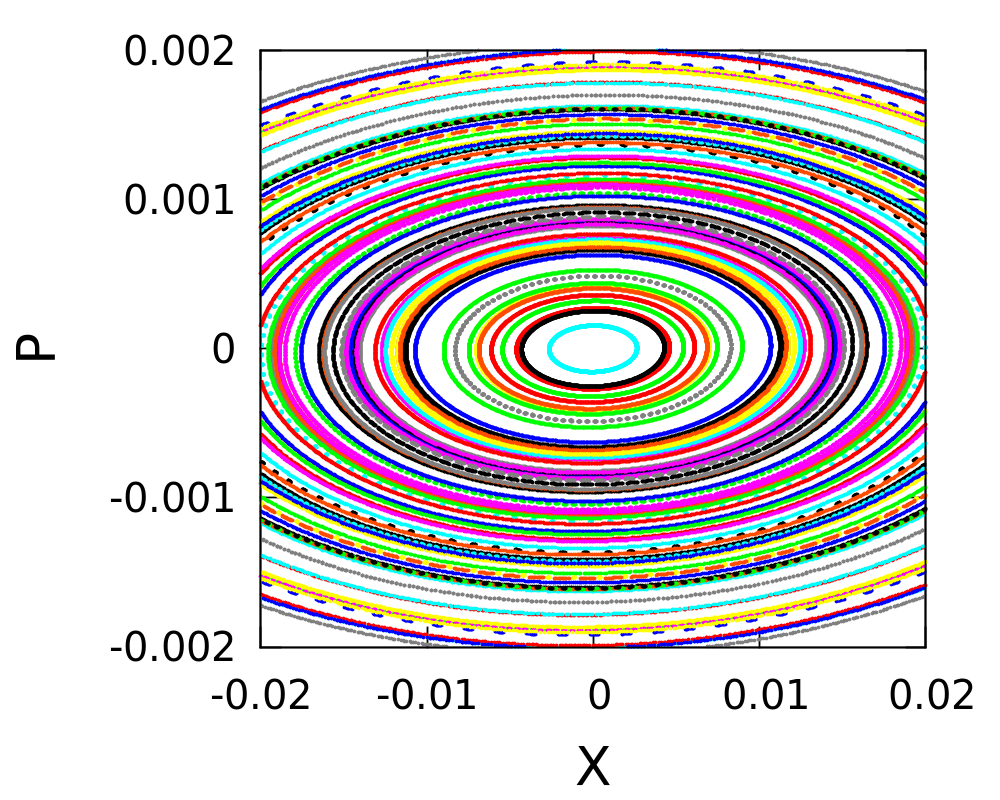}}
\subfigure[]{\includegraphics[scale=0.12]{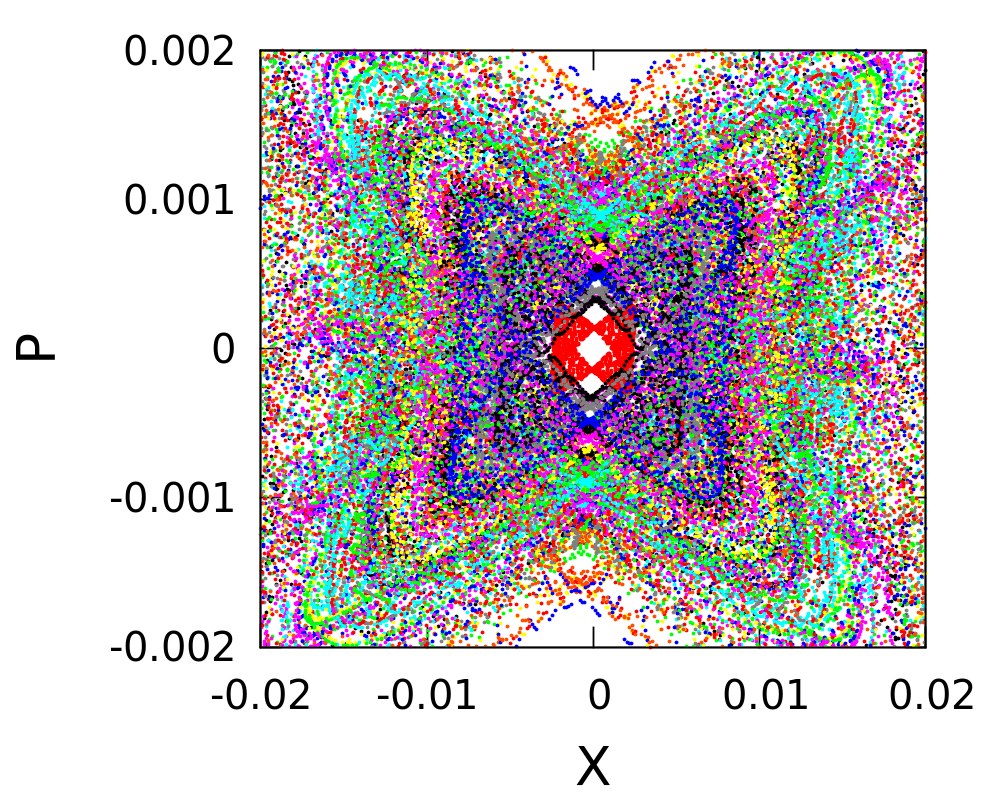}}
\subfigure[]{\includegraphics[scale=0.12]{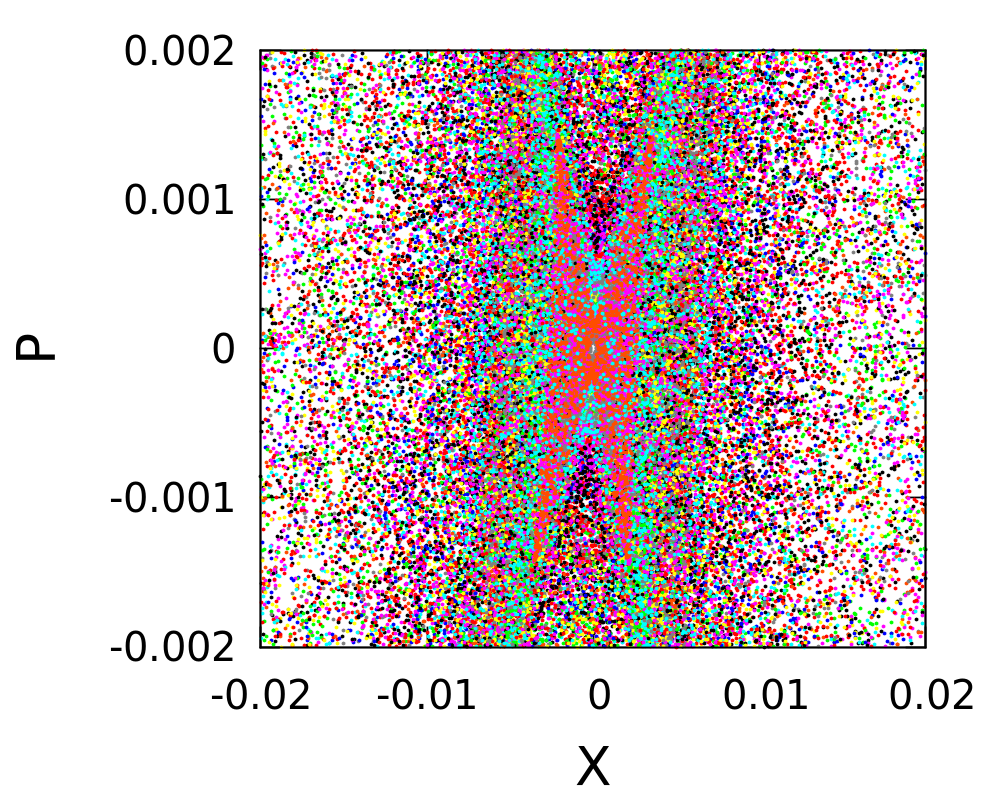}}
\subfigure[]{\includegraphics[scale=0.12]{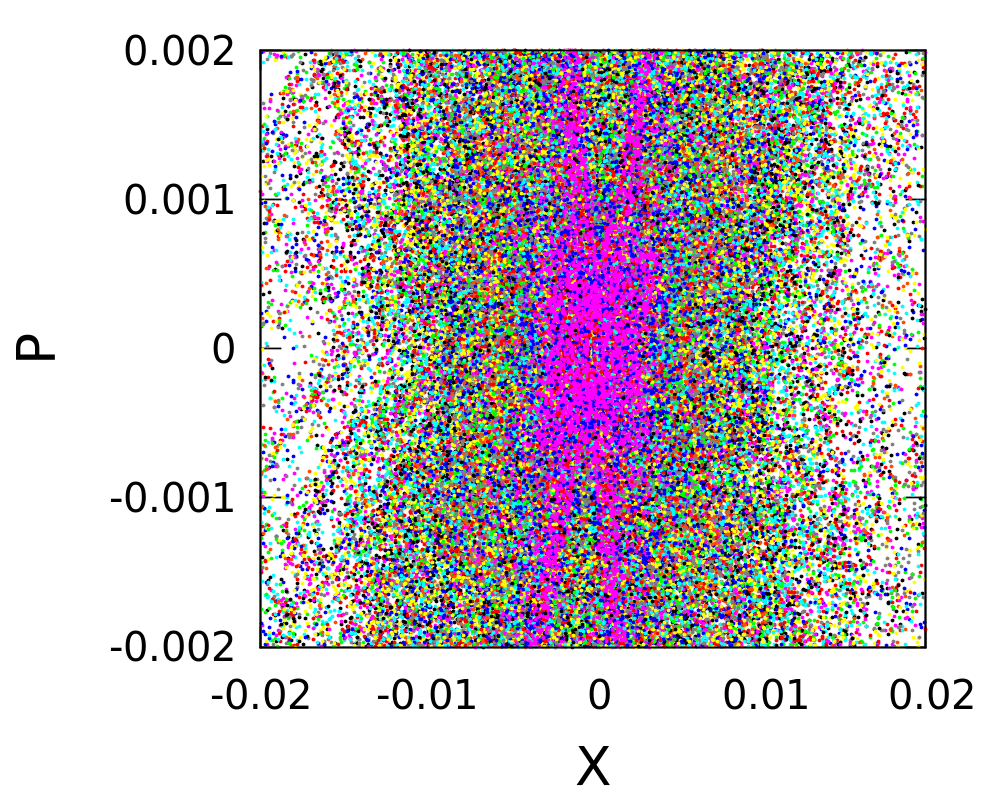}}
\caption{Phase portraits of CHM for increasing values of $\lambda$ and with interaction strength $UN_b=0.5$. Figure (a) to (d) correspond to $\lambda$=0, 0.2, 0.9, 1.8 respectively.}    
\label{fig9}
\end{figure}  
%%%%%%%%%%%%%%%%%%%%%%%%%%%%%%%%%%%%%%%%%%%%%%%%%%%%%%%%%%%%%%%%%%%%%%%%%%%%%%
where, $x_{i} = \phi_{i}$ and $p_{i} = \phi_{i} - \phi_{i-1}$. The repulsive nonlinearity gives rise to an unstable classical potential $ \sim - \frac{UN_{b}}{4}x_{i}^{4}$ due to which the classical trajectories become unstable at large values of phase space variables. To avoid this problem we choose relatively small region of phase space within which the potential is metastable, and study the effect of disorder in the phase space dynamics. In Fig.\ref{fig9}, for a fixed value of nonlinearity $UN_{b}$ we show the phase portrait with an ensemble of initial configurations for increasing values of disorder strength $\lambda$. Similar qualitative features like non interacting system are also seen in the phase space dynamics of DNLS, however the classical periodic orbits are modified due to the nonlinearity.

\subsection{Fluctuations within Bogoliubov approximation}
So far we studied the weakly interacting Bosons within mean-field approximation using macroscopic wavefunction for the condensate. It is also important to analyze the quantum fluctuations induced by the quasiperiodic disorder potential. Within Bogoliubov approximation, the quantum field operators can be approximated by, 
\begin{equation}
a_{i} = e^{-i\mu t}\left[\psi_{i} + \sum_{\nu}\left(u_{i}^{\nu}b_{\nu}e^{-i\omega_{\nu}t} + v_{i}^{\ast \nu}b_{\nu}^{\dagger}e^{i\omega_{\nu}t}\right)\right],
\label{fluctuation_op}
\end{equation}
where $\psi_{i}$ is the macroscopic wavefunction of the condensate satisfies Eq.\ref{dnlse}, $u_{i}^{\nu}$,$v_{i}^{\nu}$ are amplitudes corresponding to $\nu$th eigenmode with bosonic operators $b_{\nu}$, $b_{\nu}^{\dagger}$. Bogoliubov quasiparticle energies $\omega_{\nu}$ can be obtained from,
\begin{eqnarray}
& & -(u_{i+1} + u_{i-1}) +[\lambda \cos(2\pi \beta i) + 2 U|\psi_{i}|^{2} -\mu]u_{i}\nonumber\\
& & + U \psi_{i}^{2}v_{i}  = \omega u_{i},\\
& & -(v_{i+1} + v_{i-1}) +[\lambda \cos(2\pi \beta i) + 2 U|\psi_{i}|^{2} -\mu]v_{i} \nonumber\\
& & + U \psi_{i}^{\ast 2}u_{i} = -\omega v_{i},      
\label{bog_eqns}
\end{eqnarray}
and normalization condition gives $ \sum_{i} (u_{i}^{\nu}u_{i}^{\ast \nu'} -   v_{i}^{\nu}v_{i}^{\ast \nu'}) = \delta_{\nu \nu'}$. 
%%%%%%%%%%%%%%%%%%%%%%%%%%%%%%%%%%%%%%%%%%%%%%%%%%%%%%%%%%%%%%%%%%%%%%%%%%%%%%
\begin{figure}[ht]
\subfigure[]{\includegraphics[scale=0.48]{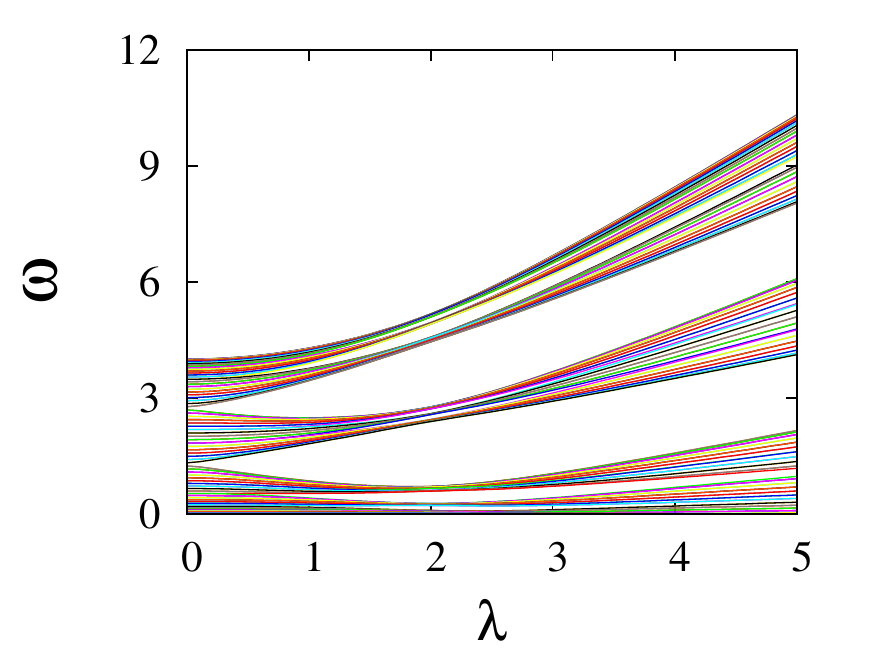}}
\subfigure[]{\includegraphics[scale=0.48]{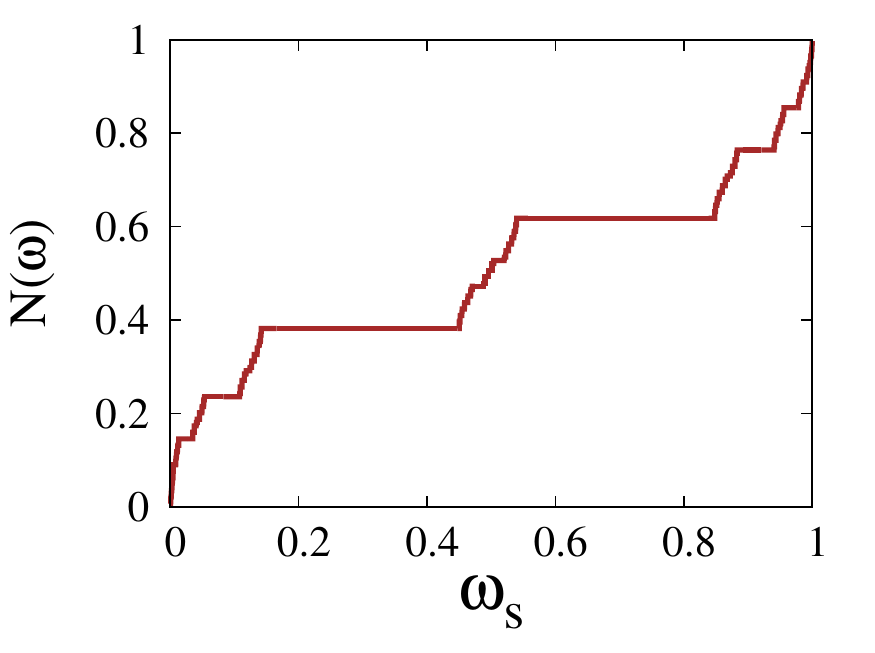}}
\subfigure[]{\includegraphics[scale=0.12]{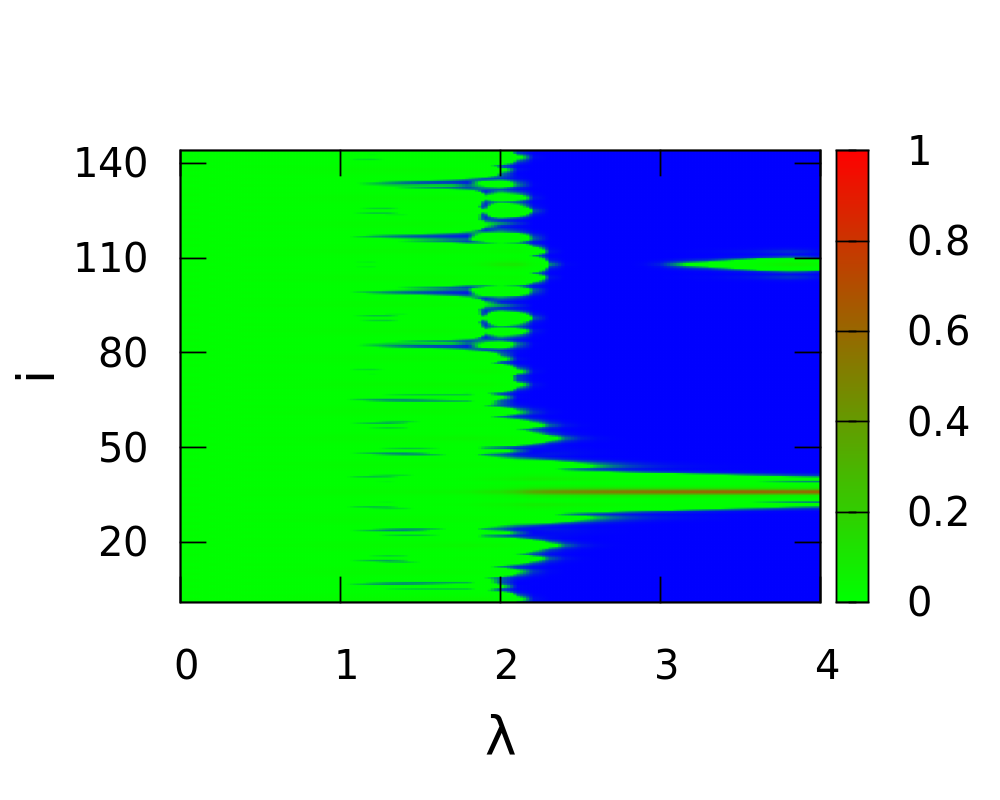}}
\subfigure[]{\includegraphics[scale=0.12]{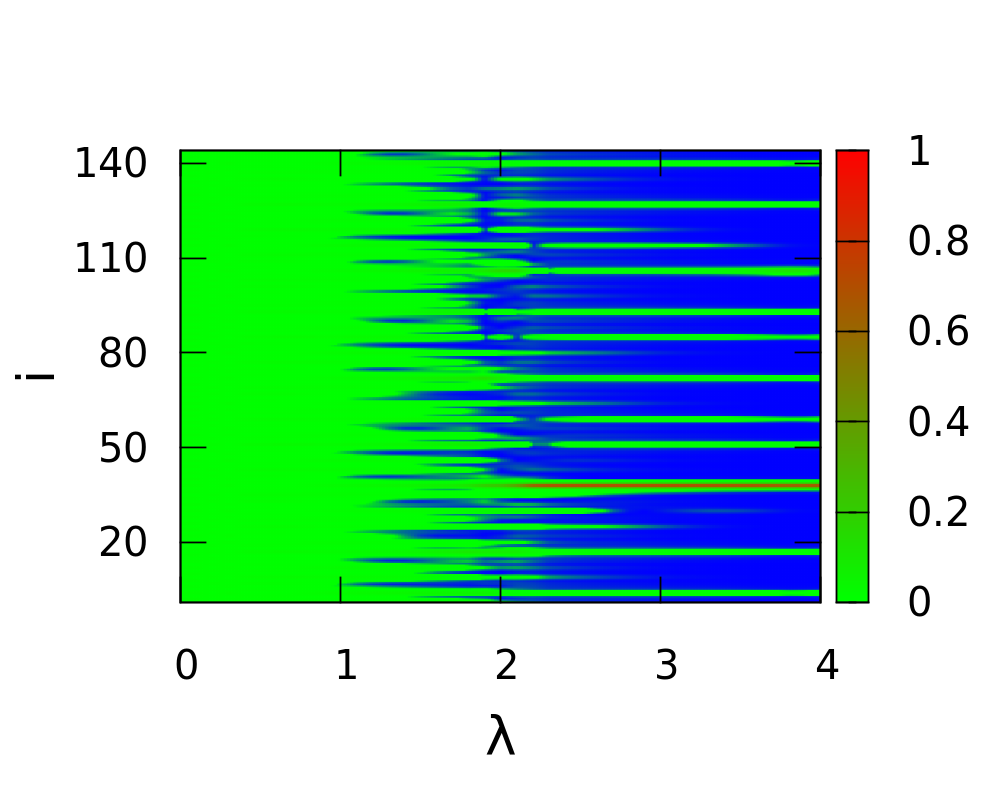}}
\caption{(a) Bogoluibov excitation spectra with increasing $\lambda$ for $UN_b=1$. (b) Devil's staircase structure of the integrated density of states $N(\omega)$ (scaled by $N(\omega_{max})$) as a function of scaled energy $\omega_{s} = (\omega - \omega_{min})/(\omega_{max} -\omega_{min})$ for $\lambda =2.5$ and $UN_b=1$ . Here $\omega_{min}$ and $\omega_{max}$ are minimum and maximum energy eigenvalues, respectively. (c)-(d) : Variation of normalized density of Bogoliubov amplitudes $|u_{\nu}|^{2}$ and $|v_{\nu}|^{2}$ respectively with $\lambda$. The index $\nu$ represents an energy state at the center of the band with $\omega_{\nu} \sim 2$.}
\label{fig10}
\end{figure}
%%%%%%%%%%%%%%%%%%%%%%%%%%%%%%%%%%%%%%%%%%%%%%%%%%%%%%%%%%%%%%%%%%%%%%%%%%%%%%

For a given interaction strength $U$ and average density of Bosons $n_{0} = N_{b}/N_{s}$, we first numerically obtain the ground state macroscopic wavefunction $\psi_{i}$, then diagonalize Eq.\ref{bog_eqns} to calculate Bogoliubov quasiparticle energies and amplitudes $u_{i}^{\nu}$, $v_{i}^{\nu}$. 
The Bogoliubov energy spectrum for increasing strength of disorder $\lambda$ is depicted in Fig.\ref{fig10}(a) for a fixed value of interaction $UN_{b}=1$ and $N_{s}=144$. The normalized integrated density of states $N(\omega)$ also shows Devil's staircase like structure as shown in Fig.\ref{fig10}(b). 
%Although the fractal nature of the integrated density of quasiparticle is 
%visible but the Devil's staircase structure (likely the fractal dimension) 
%changes with the interaction strength $Un_{0}$.   
%%%%%%%%%%%%%%%%%%%%%%%%%%%%%%%%%%%%%%%%%%%%%%%%%%%%%%%%%%%%%%%%%%%%%%%%%%%%%%%%%%%%%
\begin{figure}[ht]
\subfigure[]{\includegraphics[scale=0.48]{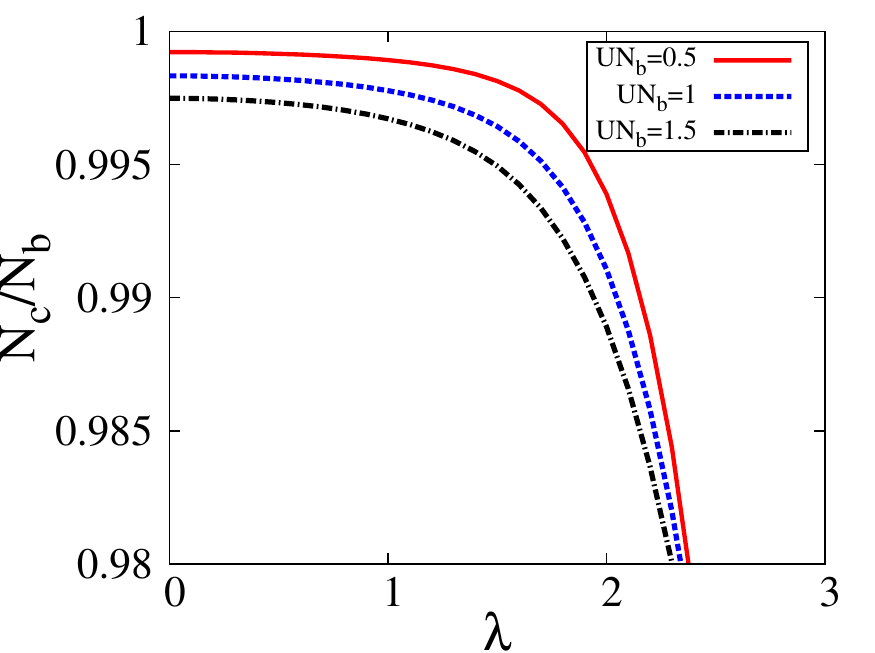}}
\subfigure[]{\includegraphics[scale=0.48]{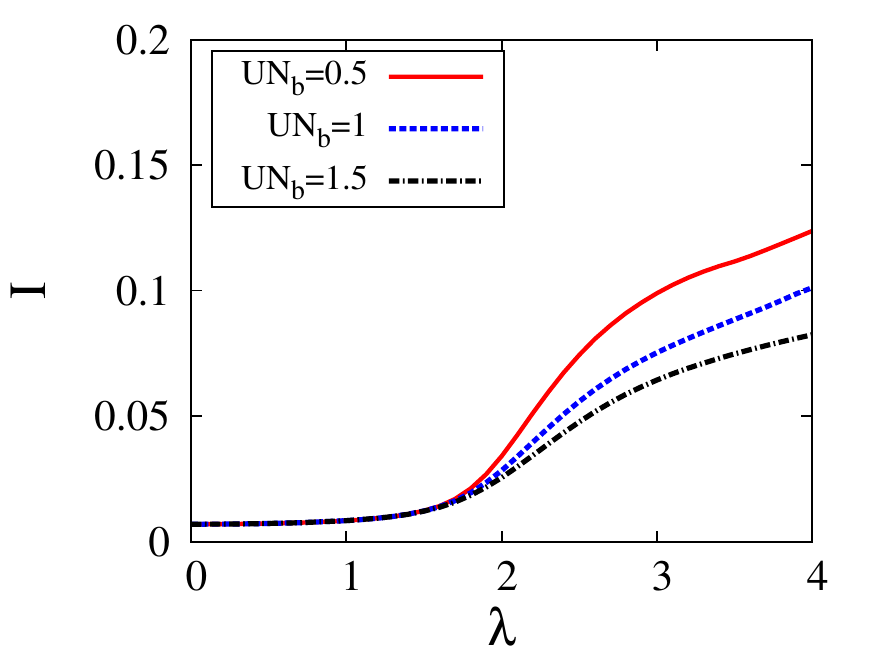}}
\subfigure[]{\includegraphics[scale=0.2]{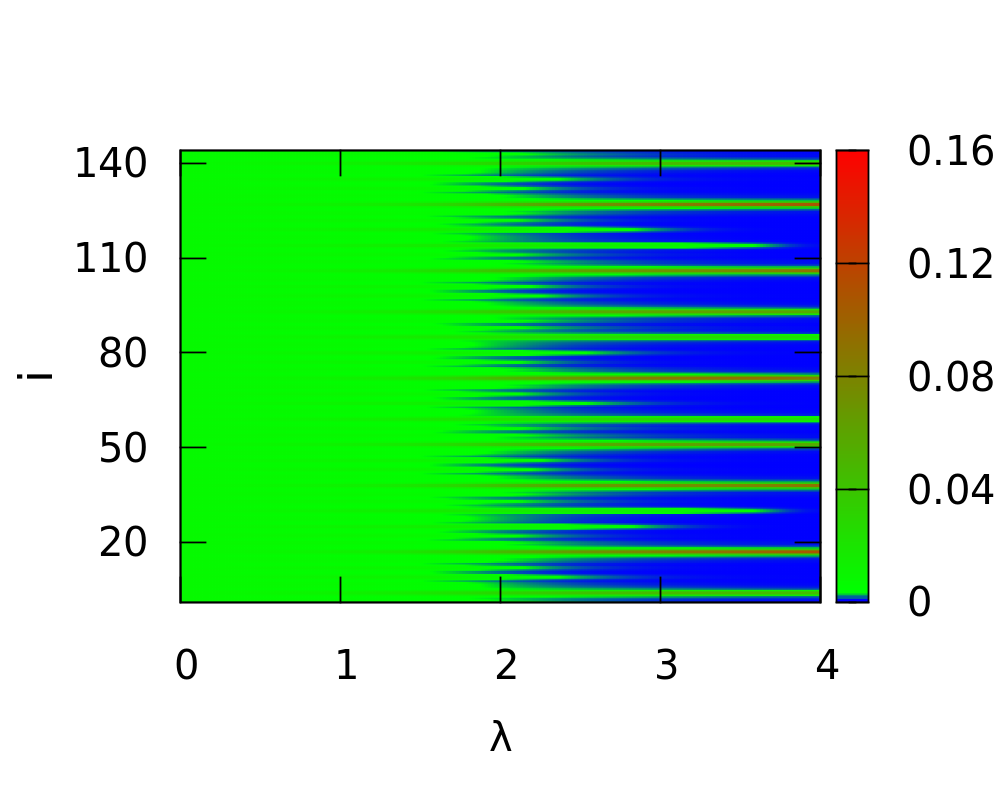}}
\caption{Variation of (a) condensate fraction $N_c/N_b$ and (b) inverse participation ratio $I$ of non-condensate density $\rho_{nc}$ with $\lambda$ for  different interaction strengths $UN_b$ as shown in the figure. (c) Spatial distribution of $\rho_{nc}$ with increasing disorder strength $\lambda$ keeping $UN_b=1$. For all the above figures, $N_s=144$ and $N_b=500$.}
\label{fig11}
\end{figure}
%%%%%%%%%%%%%%%%%%%%%%%%%%%%%%%%%%%%%%%%%%%%%%%%%%%%%%%%%%%%%%%%%%%%%%%%%%%%%%%%%%%%
Due to the quantum fluctuations, depletion of the condensate occurs and the non-condensate density $\rho_{nc}$ at zero temperature can be obtained from the Bogoliubov theory,
\begin{equation}
\rho_{nc}(i) = \sum_{\nu} |v_{i}^{\nu}|^{2}.
\label{den_nc}
\end{equation}
The condensate fraction is given by $N_{c}/N_{b} = 1 - \sum_{i,\nu}|v_{i}^{\nu}|^{2}/N_{b}$. In one dimensional system the noncondensate fraction diverges as $ \log(N_s)$, which prohibits the formation of condensate in the thermodynamic limit. However for sufficiently weak interaction a quasi condensate can form in quasi one dimensional and finite system \cite{1d_shlyap}. 
To investigate the disorder induced quantum fluctuation in a quasi-condensate in a finite lattice with $N_{s}= 144$, we calculate the condensate fraction with increasing disorder strength $\lambda$, which is shown in Fig.\ref{fig11}(a). 
In absence of disorder the quantum depletion is small in weakly interacting gas of Bosons and increases with the interaction strength. With the increase in disorder strength $\lambda$ the condensate fraction remains close to unity in the delocalized regime and then
rapidly decreases around the critical point $\lambda \approx 2$. This qualitative feature (as shown in Fig.\ref{fig11}(a)) indicates that enhanced quantum fluctuations near the localization transition can destroy the quasi condensate above $\lambda \approx 2$ and strongly correlated phases can appear. The variation of non-condensate density $\rho_{nc}$ with disorder strength $\lambda$ is depicted in Fig.\ref{fig11}(c) by color scale plot. It is interesting to note that the distinct feature of multisite localization for $\lambda > 2$ is observed even for non-condensate density. Also the IPR of normalized non-condensate density shows similar behavior as that of the condensate wavefunction and increases from $\lambda \approx 2$ (as seen from Fig.\ref{fig11}(c)). 
%This analysis clearly shows that quantum fluctuation of weakly interacting 
%Bose gas is strongly affected by the presence of quasiperiodic potential 
%particularly above the critical value $\lambda = 2$.
Localization of Bogoliubov quasiparticles and enhancement of quantum fluctuations in presence of quasiperiodic potential particularly for $\lambda \ge 2$ are clearly evident from this analysis.
%%%%%%%%%%%%%%%%%%%%%%%%%%%%%%%%%%%%%%%%%%%%%%%%%%%%%%%%%%%%%%%%%%%%%%%%%%%%%%%%%%%%
\section{Localization in the presence of a trap}
Although the localization transition in AA model occurs in thermodynamic limit, in real experimental setup a weak trapping potential is always present in order to confine the ultracold atoms. Main features of the localization can also be observed in trapped system provided the length scale of the trapping potential is larger compared to the localization length. Additionally some interesting effects due to the trap can also be seen. A harmonic trap is introduced by adding a potential $V_{i} = \frac{1}{2}\omega_{HO}^{2}(i-i_{c})^{2}$ in the Hamiltonian (Eq.\ref{Bose_hubbard}), where $\omega_{HO}$ is related to the trapping frequency, $i$ is the site index, and $i_{c}$ is the center of the trap.
%%%%%%%%%%%%%%%%%%%%%%%%%%%%%%%%%%%%%%%%%%%%%%%%%%%%%%%%%%%%%%%%%%%%%%%%%%%%%%%%%%%%%%%%%%%%%%
\begin{figure}[ht]
\subfigure[]{\includegraphics[scale=0.7]{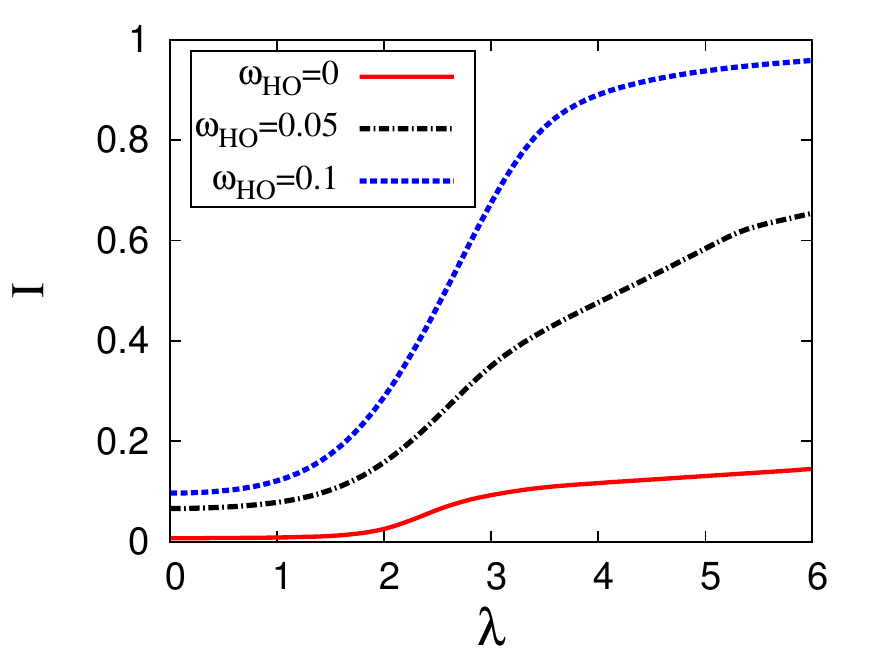}}
\subfigure[]{\includegraphics[scale=0.7]{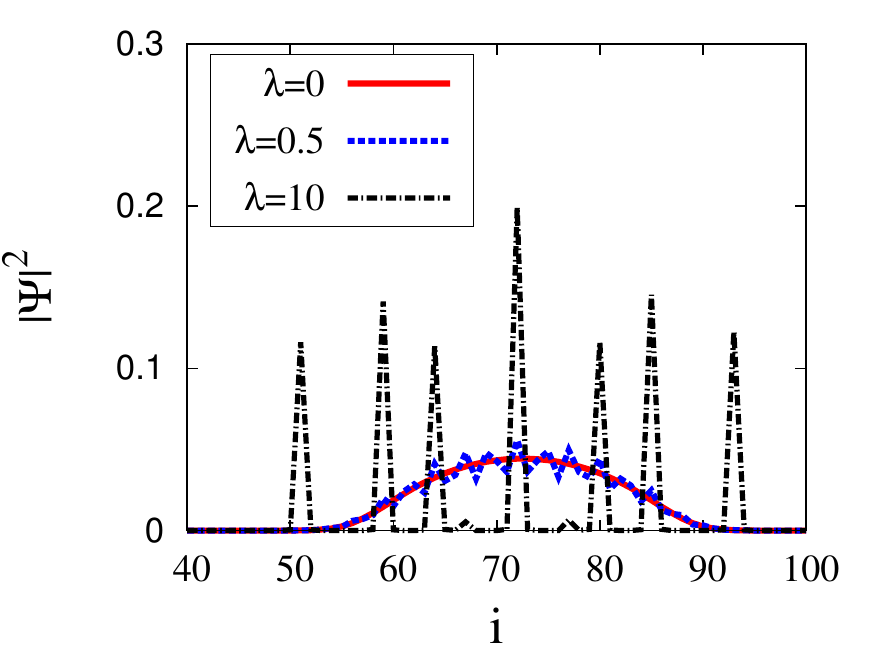}}
\caption{(a) Variation of inverse participation ratio $I$ with increasing $\lambda$ for different trap frequencies $\omega _{HO}$ (as shown in figure) keeping $UN_b=0.5$ and $N_b=500$. (b) Density distribution of harmonically trapped condensate for different $\lambda$ and for $UN_b=8$, $N_b=500$ and $\omega_{HO}=0.05$.}
\label{fig12}
\end{figure}
%%%%%%%%%%%%%%%%%%%%%%%%%%%%%%%%%%%%%%%%%%%%%%%%%%%%%%%%%%%%%%%%%%%%%%%%%%%%%%%%%%%%%%%%%%%%%%
Although in a trap the wavefunction is always localized, but in a weak trapping potential the width of the wavepacket is sharply reduced due to disorder induced localization. The density profile of the condensate in presence of a harmonic trap for different disorder strength $\lambda$ is shown in Fig.\ref{fig12}b. We calculate the IPR of the ground state wavefunction with increasing disorder strength $\lambda$ and an increase of IPR around $\lambda \approx 2$ is observed as expected. However the IPR does not vanish in the delocalized regime $\lambda <2$ and takes a small value due to the tapping potential. We have noticed earlier that in the absence of a trap the IPR increases very slowly after the localization transition due  to the repulsive interaction and the wavefunction is localized at spatially separated sites with quasi-degenerate energies. These quasi-degeneracy of onsite energies can be lifted by introducing a harmonic trap which leads to enhancement of the degree of localization which is elucidated in Fig.\ref{fig12}a where a rapid increase of IPR to unity is shown by increasing the trap frequency by a small amount.
%%%%%%%%%%%%%%%%%%%%%%%%%%%%%%%%%%%%%%%%%%%%%%%%%%%%%%%%%%%%%%%%%%%%%%%%%%%%%%%%%%%%%%%%%%%%%
\begin{figure}
\subfigure[]{\includegraphics[scale=0.9]{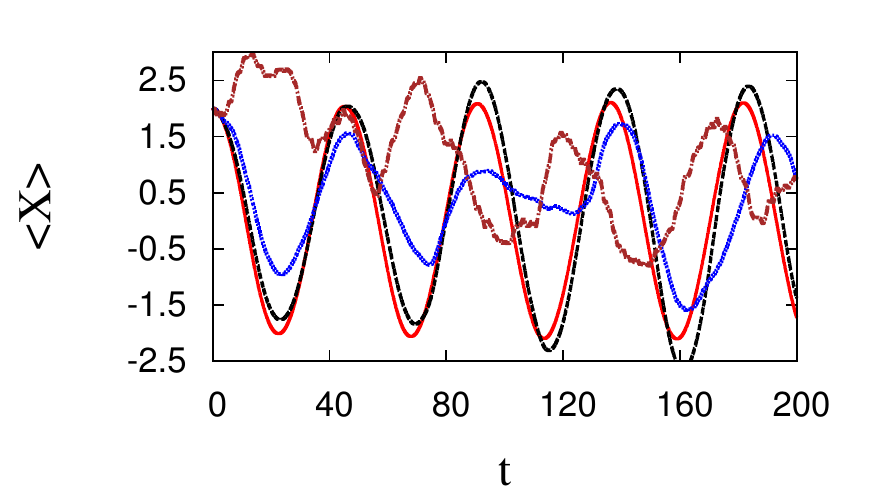}}
\subfigure[]{\includegraphics[scale=0.9]{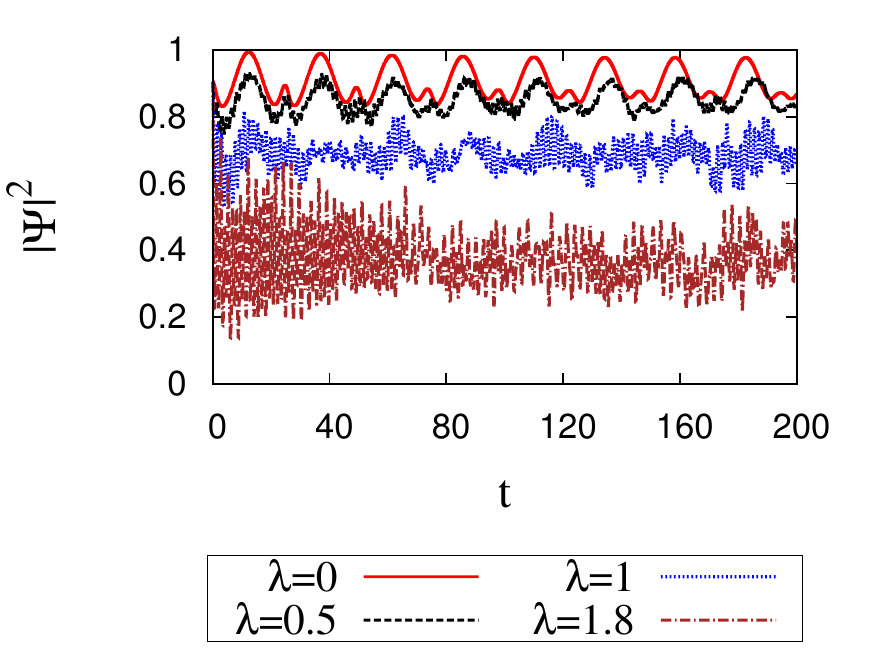}}
\caption{(a)Center of mass motion for different disorder strength $\lambda$ and for $\omega_{HO}=0.1$, $UN_b=2.5$ and $N_b=500$. For higher $\lambda$ oscillation becomes incoherent.(b)-Coherence factor which is measure of overall coherence in the system decreases for increasing $\lambda$.}
\label{fig13}
\end{figure}
%%%%%%%%%%%%%%%%%%%%%%%%%%%%%%%%%%%%%%%%%%%%%%%%%%%%%%%%%%%%%%%%%%%%%%%%%%%%%%%%%%%%%%%%%%%%%

Collective oscillation of a Bose-Einstein condensate in a trap can also show its superfluid properties. A small displacement of the condensate from the center of the trap generates center of mass (COM) oscillation  which is a well studied collective mode of Bose-Einstein condensate. Here we numerically study the motion of COM of trapped condensate in presence of quasi periodic disorder. Since the superfluidity is affected by the disorder, we calculate the coherence factor $\Psi$ of the oscillating condensate which is given by \cite{Smerzi}
\begin{equation}
\Psi = \sum_{i} \psi_{i}^{\ast} \psi_{i+1},
\label{coherence_fac}
\end{equation}
We can notice that the above expression contains information of relative phase difference between neighboring sites and  $|\Psi|^{2}$ gives a quantitative measure of overall phase coherence of the oscillating condensate.

In Fig.\ref{fig13} (a) and (b), we have shown the COM motion of the trapped condensate and coherence factor of corresponding macroscopic wavefunction for increasing disorder strength $\lambda$. It is clear that the coherence of the time dependent wavefunction decreases with increasing disorder. It is also interesting to study the variation of condensate fraction with disorder. In one dimensional trapped condensate the divergence of non-condensate fraction is less severe due to the finiteness of the trap. Like the homogeneous system, we expect enhancement of quantum fluctuation due to disorder which may destroy the condensate around a critical disorder strength $\lambda \approx 2$.
%%%%%%%%%%%%%%%%%%%%%%%%%%%%%%%%%%%%%%%%%%%%%%%%%%%%%%%%%%%%%%%%%%%%%%%%%%%%%%%%%%%%%%%%%%%%%%
\section{Conclusions}
In conclusion, we have investigated localization of both non-interacting Bosons as well as weakly interacting quasi-condensate in presence of a quasiperiodic potential.
Apart from calculating various physical properties, understanding localization transition through the approach of classical Hamiltonian map is one of the main result of this work.

In the non-interacting Aubry-Andr\'e model the localization transition can be identified by the vanishing `superfluid fraction' and the rise of IPR at the critical strength of quasiperiodic potential $\lambda=2$. A classical Hamiltonian map is constructed from the Schr\"odinger equation. The phase space trajectories of the  corresponding classical map shows periodic orbits for small disorder strength. With the increasing disorder strength $\lambda$ chaotic behavior is observed at the outer region of the phase space and finally near the critical value $\lambda = 2$ all periodic orbits are destroyed due to the onset of chaos in the localized regime. The parameter $\beta=(\sqrt{5}-1)/2$ is chosen to be an irrational Diophantine number(inverse of golden mean) which is an essential requirement for localization transition. This has been elucidated by means of Hamiltonian map for $\lambda > 2$ where the destruction of periodic orbits by successive rational approximation of $\beta$ indicates onset of localization. 

We have investigated the localization of weakly interacting Bose gas in quasiperiodic potential by mean-field approach. Unlike non-interacting case, the vanishing of SFF and rise of IPR does not take place at same strength of disorder $\lambda$. Due to the repulsive interactions, the macroscopic wavefunction is localized at many sites with quasi-degenerate energies for $\lambda >2$. The multisite localization of the interacting system is manifested by a much slower increase of IPR starting from $\lambda =2$. With increasing the strength of quasiperiodic potential, the number of  sites over which the condensate is localized decreases and finally the wavefunction becomes localized at a single site for very large value of $\lambda$ and IPR approaches to unity. The SFF decreases with increasing disorder strength $\lambda$ and vanishes at $\lambda>2$ due to the repulsive interaction. The repulsive interaction gives rise to an unstable nonlinear potential in the CHM approach, due to which the stable region of phase space decreases. In the phase portrait the stable region containing the periodic orbits decreases with increasing $\lambda$ and finally the onset of chaos at $\lambda \approx 2$ signifies localization of the wavefunction. Further, the Bogoliubov quasiparticle spectrum has been calculated numerically. We notice that disorder enhances the quantum fluctuations due to which the condensate fraction of a finite system decreases rapidly around $\lambda \approx 2$. This indicates the possible formation of glassy phase and multisite localized insulators.

 %which shows a remarkable spectral property. The scaled integrated density of states of the quasi-particle energies shows devil's staircase like fractal structure which becomes universal for any interaction strength at a critical value of disorder strength $\lambda = 2$.

Finally we also considered the effect of trapping potential on the localization transition. In the localized regime, the number of sites over which the wavefunction is localized reduces due to the presence of a trap which has been shown from the rapid increase in the IPR by tuning the trap frequency. The center of mass motion of the condensate in a harmonic trap also shows the signature of localization. The center of mass oscillations become incoherent with increasing disorder. To summarize, the present study provides a clear picture of localization of non-interacting and weakly interacting Bose gas in the presence of a quasiperiodic disorder and it reveals various interesting features which are interesting for both academic point of view, as well for future experiments.

\end{document}